\documentclass[final]{arxiv}

\usepackage{graphicx}
\usepackage{natbib}
\renewcommand{\arraystretch}{0.5}  

\begin{document}

\addtolength{\voffset}{0.5in}  

\doublespace

\title{\LARGE{Binary Asteroid Systems:  Tidal End States\\and Estimates of Material Properties}}
\author{\Large{Patrick A. Taylor$^{1}$ and Jean-Luc Margot$^{2}$}}
\affil{$^{1}$Arecibo Observatory, $^{2}$UCLA}
\email{ptaylor@naic.edu}

\journame{Icarus}
\submitted{29 June 2010}
\revised{17 January 2011}
\accepted{18 January 2011}
\pubonline{1 February 2011}
\pubprint{April 2011 in Volume 212, pp. 661--676}

\pages{39}
\tables{3}
\figures{6}

\quad\newline

\singlespace  

\noindent
NOTICE: this is the author's version of a work that was accepted for publication in Icarus. Changes resulting 
from the publishing process, such as peer review, editing, corrections, structural formatting, and other 
quality control mechanisms may not be reflected in this document. Changes may have been made to this work 
since it was submitted for publication. A definitive version was subsequently published in Icarus 212, 
661--676, April 2011, DOI: 10.1016/j.icarus.2011.01.030.\\
\\
Publisher's copy:  http://dx.doi.org/10.1016/j.icarus.2011.01.030 

\quad\clearpage

\addtolength{\voffset}{-0.5in}


\noindent ABSTRACT:\\

The locations of the fully despun, double synchronous end states of tidal evolution, where the rotation rates 
of both the primary and secondary components in a binary system synchronize with the mean motion about the 
center of mass, are derived for spherical components.  For a given amount of scaled angular momentum 
$J/J^{\prime}$, the tidal end states are over-plotted on a tidal evolution diagram in terms of mass ratio of 
the system and the component separation (semimajor axis in units of primary radii).  Fully synchronous orbits 
may not exist for every combination of mass ratio and angular momentum; for example, equal-mass binary systems 
require $J/J^{\prime} > 0.44$.  When fully synchronous orbits exist for prograde systems, tidal evolution 
naturally expands the orbit to the stable outer synchronous solution.  The location of the unstable inner 
synchronous orbit is typically within two primary radii and often within the radius of the primary itself.  
With the exception of nearly equal-mass binaries, binary asteroid systems are in the midst of lengthy tidal 
evolutions, far from their fully synchronous tidal end states.  Of those systems with unequal-mass components, 
few have even reached the stability limit that splits the fully synchronous orbit curves into unstable inner 
and stable outer solutions.

Calculations of material strength based on limiting the tidal evolution time to the age of the Solar System 
indicate that binary asteroids in the main belt with 100-km-scale primary components are consistent with being 
made of monolithic or fractured rock as expected for binaries likely formed from sub-catastrophic impacts in 
the early Solar System.  To tidally evolve in their dynamical lifetime, near-Earth binaries with km-scale 
primaries or smaller created via a spin-up mechanism must be much weaker mechanically than their main-belt 
counterparts even if formed in the main belt prior to injection into the near-Earth region.  Small main-belt 
binaries, those having primary components less than 10 km in diameter, could bridge the gap between the large 
main-belt binaries and the near-Earth binaries, as, depending on the age of the systems, small main-belt binaries 
could either be as strong as the large main-belt binaries or as weak as the near-Earth binaries.  The inherent 
uncertainty in the age of a binary system is the leading source of error in calculation of material properties, 
capable of affecting the product of rigidity $\mu$ and tidal dissipation function $Q$ by orders of magnitude.  
Several other issues affecting the calculation of $\mu Q$ are considered, though these typically affect the 
calculation by no more than a factor of two.  We also find indirect evidence within all three groups of binary 
asteroids that the semimajor axis of the mutual orbit in a binary system may evolve via another mechanism (or 
mechanisms) in addition to tides with the binary YORP effect being a likely candidate.\\
\\

\pagebreak

\section{Introduction}
\label{sec:intro}

Over 20 years ago,~\citet{weid89} asked, ``Do asteroids have satellites?''  Today, binary systems have been 
discovered in every dynamical class of small Solar System bodies from near-Earth asteroids to Mars-crossers and 
main-belt asteroids, among the Jupiter Trojans, and in the outer Solar System in the Kuiper belt.  Beginning 
with the Galileo spacecraft's serendipitous discovery in 1993 of tiny satellite Dactyl orbiting (243) 
Ida~\citep{chap95,belt96} while on its cruise to Jupiter and continuing with the success of radar, lightcurve 
photometry, ground-based adaptive-optics imaging, and Hubble Space Telescope imaging, reviewed 
by~\citet{merl02},~\citet{rich06}, and~\citet{noll08}, over 180 small Solar System bodies are suspected to be 
binary or multiple systems.\footnote{The value of more than 180 suspected binary or multiple systems is taken 
from http://www.johnstonsarchive.net/astro/asteroidmoons.html and based on references therein.  Lists of binary 
and multiple asteroid parameters are available from the Ondrejov Asteroid Photometry Project 
(http://www.asu.cas.cz/$\sim$asteroid/binastdata.htm) and the Planetary Data System 
(http://sbn.psi.edu/pds/asteroid/EAR$\_$A$\_$COMPIL$\_$5$\_$BINMP$\_$V3$\_$0/data/).}  In this work, we build 
upon the ideas presented in~\citet{weid89} to illustrate the tidal end states of binary asteroid systems and 
discuss the inherent material strength of the bodies implied by tidal evolution.

The Keplerian orbit of the two components in a binary system about the center of mass allows one to determine 
the mass of the system and, with an estimate of the component sizes, the (assumed equal) densities of the 
bodies.  Density estimates combined with taxonomic classifications hint at the internal structure of 
asteroids~\citep{brit02}.  When compared to analog meteorites with known bulk densities, low densities found 
for rocky asteroids imply that, rather than monolithic bodies, asteroids are aggregate bodies made up of 
many fragments with varying degrees of porosity.  Aggregates can range from low-porosity fractured or shattered 
bodies to porous, cohesionless rubble piles with ``gravitational aggregate'' acting as a catch-all term for 
bodies comprised of many fragments, independent of porosity, that are held together by their collective 
gravity~\citep{rich02}.  The tidal evolution of these binary systems is intimately tied to the internal structure 
and material strength of the bodies involved, and we will exploit this dependence to estimate the combined 
effect of rigidity and energy dissipation in the bodies to determine if asteroids tend to have solid, fractured, 
or shattered interiors.  

Hints of the internal structure of asteroids also come from the most probable binary formation mechanisms for 
each population:  near-Earth binaries likely form through rotational disruption via spin-up by thermal torques 
due to sunlight~\citep[the YORP effect;][]{bott02,wals08yorp} or close planetary encounters~\citep{rich98,wals06}, 
1- to 10-km-scale binaries in the main belt may also form through YORP spin-up~\citep{prav07}, 100-km-scale 
main-belt binaries likely form through sub-catastrophic impacts~\citep[\textit{e.g.,}][]{durd04}, and Kuiper-belt 
binaries may have formed very early on via gravitational collapse~\citep{nesv10} or later on via collisions or a 
flavor of dynamical capture~\citep[see][for a review]{noll08}.  Binary formation through rotational disruption at 
the very least implies a fractured interior, and if satellite formation is due to the accretion of smaller particles 
spun off the parent body, as modeled by~\citet{wals08yorp}, part or all of the parent body is likely an aggregate 
of smaller pieces and structurally weak.  For sub-catastrophic impacts, the satellite is either a solid shard or an 
aggregate made up of impact ejecta while the parent body may remain solid or fractured rather than shattered.  Thus, 
we anticipate that examination of tidal evolution in binary asteroid systems will find that large main-belt binaries 
are consistent with solid or fractured rock, while near-Earth and small main-belt binaries are consistent with 
structurally weaker, shattered or porous gravitational aggregates.  

Section~\ref{sec:angmom} introduces the important dynamical quantities of binary systems, angular momentum and 
energy.  In Section~\ref{sec:synch}, we derive and illustrate for a range of angular momenta the fully despun, 
double synchronous orbits that are end states of tidal evolution and then discuss stability limits and energy 
regimes in Section~\ref{sec:stablimit}.  Using the known properties of binary asteroids in the near-Earth region 
and the main belt, in Section~\ref{sec:material} we constrain the material properties of the bodies based on 
tidal evolution.


\section{Angular momentum and energy content}
\label{sec:angmom}

The dynamics of a binary system with primary of mass $M_{\rm p}$ and secondary of mass $M_{\rm s}$ in a 
mutual orbit about their common center of mass can be described equivalently by a system where a body of 
mass $M_{\rm p}M_{\rm s}/(M_{\rm p}+M_{\rm s})$ orbits a stationary mass, $M_{\rm p}+M_{\rm s}$.  The 
orbital angular momentum of such a system is 
$L=M_{\rm p}M_{\rm s}/(M_{\rm p}+M_{\rm s})\sqrt{G\,(M_{\rm p}+M_{\rm s})\,a\,(1-e^{2})}$ where $G$ is the 
gravitational constant and $a$ and $e$ are the semimajor axis and eccentricity of the mutual orbit, 
respectively.  Defining the component mass ratio 
$q=M_{\rm s}/M_{\rm p}=(\rho_{\rm s}/\rho_{\rm p})(R_{\rm s}/R_{\rm p})^3\le1$, where $\rho$ and $R$ are 
the bulk density and radius, respectively, of the components, using the system mass 
$M_{\rm sys}=M_{\rm p}+M_{\rm s}=(1+q)\,M_{\rm p}$, and applying Kepler's Third Law, 
$n^2a^3=G\left(M_{\rm p}+M_{\rm s}\right)$ in terms of the mean motion $n$, the orbital angular momentum
is:

\begin{equation}
L~=~\frac{q}{\left(1+q\right)^2}\,M_{\rm sys}\,a^2\,n\left(1-e^2\right)^{1/2}~=~\frac{q}{1+q}\,M_{\rm p}\,a^2\,n\left(1-e^2\right)^{1/2}.
\label{eq:Lkepler}
\end{equation}

\noindent
For principal axis rotation, the spin angular momentum of the system is given by the moments of inertia 
$I=\alpha MR^{2}$ of the components and their spin rates $\omega$ as:

\begin{equation}
S~=~I_{\rm p}\,\omega_{\rm p}~+~I_{\rm s}\,\omega_{\rm s}~=~\alpha_{\rm p}M_{\rm p}R_{\rm p}^{2}\,\omega_{\rm p}\left(1+\frac{\alpha_{\rm s}}{\alpha_{\rm p}}\left(\frac{\rho_{\rm p}}{\rho_{\rm s}}\right)^{2/3}q^{5/3}\,\frac{\omega_{\rm s}}{\omega_{\rm p}}\right).
\label{eq:S}
\end{equation}

\noindent
The coefficient $\alpha$ is 2/5 for a uniform-density, rotating sphere, but varies with the shape of the body 
and the density profile of the interior.  

For comparison, suppose a rapidly spinning, uniform-density parent body sheds mass in such a way to conserve 
angular momentum and produce the aforementioned mutually orbiting binary system.  When a spherical body with 
mass $M$ and radius $R$ spins at the breakup rate $\omega_{\rm break}$ without cohesion among its constituent 
particles, the inward acceleration due to gravity $GM/R^2$ at the equator is balanced by the outward centrifugal 
acceleration $\omega_{\rm break}^2R$ due to rotation such that $\omega_{\rm break}=\sqrt{GM/R^3}$.  The angular 
momentum $J$ contained in the critically rotating sphere is $I\omega_{\rm break}$ or: 

\begin{equation}
J~=~\frac{2}{5}\,\sqrt{GM_{\rm sys}^3 R_{\rm eff}},
\label{eq:Jparent}
\end{equation}

\noindent
where the mass and radius of the parent body have been written as the total mass and effective radius of the
binary system, 
$R_{\rm eff}=(R^{3}_{\rm p}+R^{3}_{\rm s})^{1/3}=\left(1+\frac{\rho_{\rm p}}{\rho_{\rm s}}\,q\right)^{1/3}R_{\rm p}$,
produced by the breakup of the parent body.  Subsequently, the total angular momentum of a 
binary system $J=L+S$ is often normalized by 
$J'=\sqrt{GM^{3}_{\rm sys}R_{\rm eff}}$ (attributed to~\citet{darw87}) such that $J/J'\sim0.4$ indicates 
the binary could have formed by mass shedding from the spin-up of a single spherical strengthless parent body.  
\citet{prav07} normalize the total angular momentum of the binary system by that of the critically rotating 
spherical parent body (Eq.~\ref{eq:Jparent}) such that their scaling parameter $\alpha_{\rm L}=(J/J')/0.4$, 
and $\alpha_{\rm L}=1$ has the same implication for binary formation.  We utilize the $J^{\prime}$ 
normalization throughout this work.

The minimum and maximum separations of two components in a binary system are limited by the physical 
size of the components and the total angular momentum of the system, respectively.  The separation of 
two components is naturally bounded from below by the contact condition where, at a separation of 
$R_{\rm p}+R_{\rm s}$, the components are resting against one another.  In terms of the semimajor axis 
and the radius of the primary, the contact limit is:

\begin{equation}
\left(\frac{a}{R_{\rm p}}\right)_{\rm min}~=~1+\frac{R_{\rm s}}{R_{\rm p}}~=~1+\left(\frac{\rho_{\rm p}}{\rho_{\rm s}}\,q\right)^{1/3}.
\label{eq:contact}
\end{equation}

\noindent
The contact limit ranges from $a/R_{\rm p}=1$ to $2$, akin to a pea resting on the surface of a basketball 
to two basketballs in contact.  One could consider the Roche limit as a soft lower bound on the separation 
of the bodies, as opposed to the hard limit of physical contact, but with a modest amount of cohesion, a 
secondary may exist within the Roche limit and down to the contact limit without disrupting~\citep{hols08,tayl10cmda}.  
An upper bound is placed on the separation of the components by the total angular momentum content of the 
system.  If the entire budget of angular momentum 
$\displaystyle{J=\left(J/J^{\prime}\right)\sqrt{GM_{\rm sys}^3R_{\rm eff}}}$ is transferred to the circular 
mutual orbit of the components, which are no longer spinning ($S\rightarrow0$ in Eq.~\ref{eq:S}), the 
maximum attainable separation according to Eq.~\ref{eq:Lkepler} with $e=0$ is:

\begin{equation}
\left(\frac{a}{R_{\rm p}}\right)_{\rm max}~=~\left(J/J^{\prime}\right)^2\,\frac{\left(1+q\right)^4}{q^2}\left(1+\frac{\rho_{\rm p}}{\rho_{\rm s}}\,q\right)^{1/3},
\label{eq:angmomlimit}
\end{equation}

\noindent
which increases quickly with decreasing $q$, allowing for binaries with smaller secondaries to have
much wider separations than binaries with similar-size components for the same amount of angular momentum.

Ignoring external influences on the system, the total energy of the binary $E$ is comprised of the rotations 
of the components $\displaystyle{\left(1/2\right)I\omega^{2}}$, the orbital motion about the center of mass 
$\displaystyle{\left(1/2\right)M_{\rm p}M_{\rm s}/\left(M_{\rm p}+M_{\rm s}\right)a^{2}n^{2}}$, and the mutual 
gravitational potential for a two-sphere system $\displaystyle{-GM_{\rm p}M_{\rm s}/a}$ such that:

\begin{equation}
E~=~\frac{1}{2}I_{\rm p}\omega_{\rm p}^{2}~+~\frac{1}{2}I_{\rm s}\omega_{\rm s}^{2}~-~\frac{1}{2}\frac{M_{\rm p}M_{\rm s}}{M_{\rm p}+M_{\rm s}}a^{2}n^{2}.
\label{eq:Etot}
\end{equation}

\noindent
While shape deformation may occur due to changes in spin rate~\citep{harr09,hols10} as the binary system evolves,  
we assume that this does not result in significant interchange of self-potential energy with the terms in 
Eq.~\ref{eq:Etot}.  The energy accounted for in Eq.~\ref{eq:Etot} is free to be traded among its components via 
the dynamical interaction of the two bodies, and the total energy can be dissipated as heat as a result of 
internal friction due to tidal flexure.  Depending on the amount of rotational and translational energy of the 
system compared to the mutual gravitational potential energy, the total energy may either be positive or 
negative.  A negative total energy requires the system to remain bound, while a positive total energy could 
allow the components to unbind themselves through their dynamical interaction.


\section{Fully synchronous orbits}
\label{sec:synch}

The natural evolution of a binary system is through mutual tidal interaction.  The differential gravity
across each body due to the proximity of its companion acts to evolve the rotation states of the individual
bodies as well as their physical separation; tides tend to evolve rapidly rotating components in close 
proximity to a more widely separated system with more slowly rotating components.  The end state of tidal 
evolution for a binary system is a fully despun, double synchronous orbit (hereafter called a fully 
synchronous orbit), where the spin rates of both components, $\omega_{\rm p}$ and $\omega_{\rm s}$, equal the 
mean motion in the mutual orbit $n$.  The classic example of such an end state is in the Pluto-Charon 
system where the bodies are face-locked~\citep{chri78}, meaning they keep the same sides facing one another 
because the periods of rotation and revolution have synchronized. 

The scaled angular momentum of the system $J/J^{\prime}$ at any time is given by the sum of 
Eq.~\ref{eq:Lkepler} and Eq.~\ref{eq:S} divided by $J^{\prime}$.  Assuming a circular mutual orbit, which the 
majority of binary asteroid systems have, and after setting the spin rates $\omega_{\rm p}$ and $\omega_{\rm s}$ 
equal to the mean motion $n$, replacing $n$ with a function of the semimajor axis $a$ via Kepler's Third 
Law,\footnote{In using $n^{2}a^{3}=G\left(M_{\rm p}+M_{\rm s}\right)$ here, we imply the components of the 
binary system have spherical shapes or may be treated as point masses.} and some rearranging, the locations 
of the fully synchronous orbits $a_{\rm sync}/R_{\rm p}$ are the solutions to the quasi-quadratic equation: 

\begin{equation}
\frac{1}{\alpha_{\rm p}}\,\frac{q}{1+q}\,\frac{1}{1+\frac{\alpha_{\rm s}}{\alpha_{\rm p}}\left(\frac{\rho_{\rm p}}{\rho_{\rm s}}\right)^{2/3}q^{5/3}}\left(\frac{a_{\rm sync}}{R_{\rm p}}\right)^{2}~-~\frac{J/J^{\prime}}{\alpha_{\rm p}}\,\frac{\left(1+\frac{\rho_{\rm p}}{\rho_{\rm s}}q\right)^{1/6}\left(1+q\right)}{1+\frac{\alpha_{\rm s}}{\alpha_{\rm p}}\left(\frac{\rho_{\rm p}}{\rho_{\rm s}}\right)^{2/3}q^{5/3}}\left(\frac{a_{\rm sync}}{R_{\rm p}}\right)^{3/2}~+~1~=~0.
\label{eq:synch}
\end{equation}

\noindent
If the components are spherical and have similar uniform densities, the condition for a fully synchronous 
orbit reduces to:

\begin{equation}
\frac{5}{2}\,\frac{q}{1+q}\,\frac{1}{1+q^{5/3}}\left(\frac{a_{\rm sync}}{R_{\rm p}}\right)^{2}~-~\frac{5}{2}\,\frac{\left(1+q\right)^{7/6}}{1+q^{5/3}}\left(J/J^{\prime}\right)\left(\frac{a_{\rm sync}}{R_{\rm p}}\right)^{3/2}~+~1~=~0.
\label{eq:synchsphere}
\end{equation}

\begin{figure}[!p]
\begin{center}
\includegraphics[angle=90., scale=0.60]{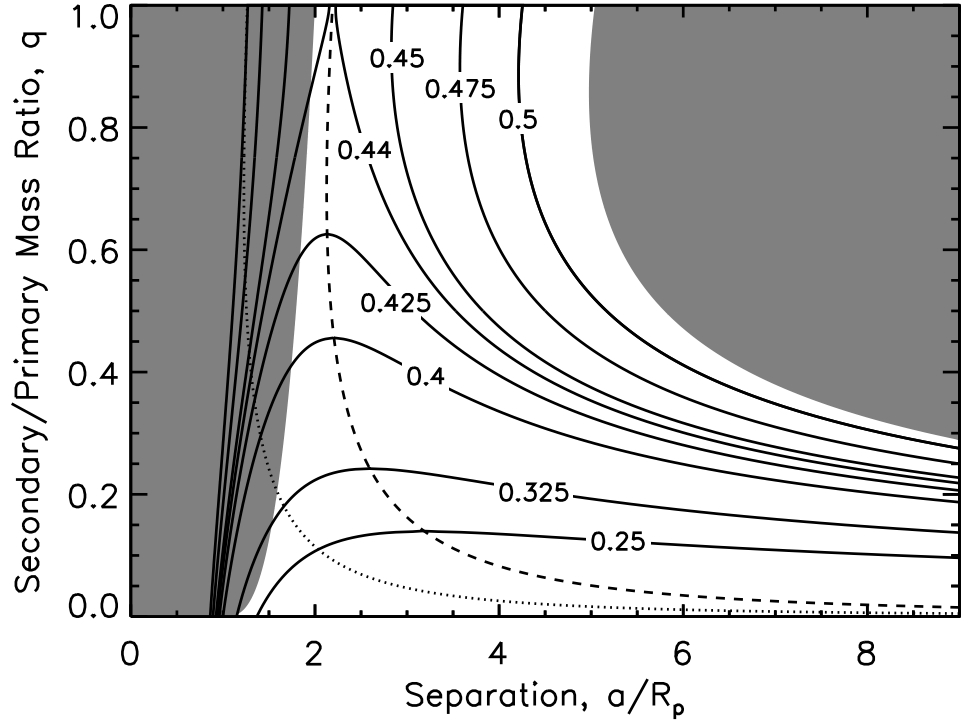}
\caption[]{Component separations $a_{\rm sync}/R_{\rm p}$ for the fully synchronous orbits of binary 
systems of given a mass ratio $q$ and scaled angular momentum $J/J^{\prime}$.  Solid lines indicate the 
solutions to Eq.~\ref{eq:synchsphere} for 
$J/J^{\prime}=0.25,\,\,0.325,\,\,0.4,\,\,0.425,\,\,0.44,\,\,0.45,\,\,0.475,$ and $0.5$.  The shaded regions 
indicate separations that are inaccessible to the binary components.  The region at left 
(Eq.~\ref{eq:contact}) requires the components to be in contact, while the region at right 
(Eq.~\ref{eq:angmomlimit}; shown only for $J/J^{\prime}=0.5$) is disallowed by angular momentum conservation
 for circular orbits.  The dashed line indicates the synchronous stability limit 
(see Section~\ref{sec:stablimit}) that divides the two solutions for each mass ratio, if they exist, into 
an unstable inner synchronous orbit and a stable outer synchronous orbit.  In most cases, the unstable 
inner synchronous orbit is within the contact limit.  The dotted line indicates where fully synchronous 
systems have total energy $E=0$ (also see Section~\ref{sec:stablimit}).}
\label{fig:synch}
\end{center}
\end{figure}

\noindent
Depending on the total angular momentum of the system $J/J^{\prime}$ and the mass ratio $q$, the equation
above may have two, one, or zero real solutions corresponding to inner and outer fully synchronous orbits, a 
single degenerate fully synchronous orbit, or the absence of a valid fully synchronous orbit, respectively, 
for the system.  Solutions to Eq.~\ref{eq:synchsphere} for all $q$ are shown in Fig.~\ref{fig:synch} for 
various values of $J/J^{\prime}$ where the number of solutions for a specific $q$ is given by the number of 
intersections between the curve of the relevant $J/J^{\prime}$ value for the system and the horizontal line 
of constant mass ratio $q$ that the system will tidally evolve along.  From Fig.~\ref{fig:synch}, binary 
systems with smaller secondaries can clearly attain much wider separations than binary systems with components 
of similar size as predicted from the angular momentum limit in Eq.~\ref{eq:angmomlimit}.  Larger amounts of 
angular momentum are required to support systems with larger secondaries, and as angular momentum increases, 
the synchronous orbit curves surround one another with the inner synchronous orbit moving inward while the 
outer synchronous orbit moves outward approaching the angular momentum limit for smaller mass ratios.


\begin{figure}[!p]
\begin{center}
\includegraphics[angle=90., scale=0.60]{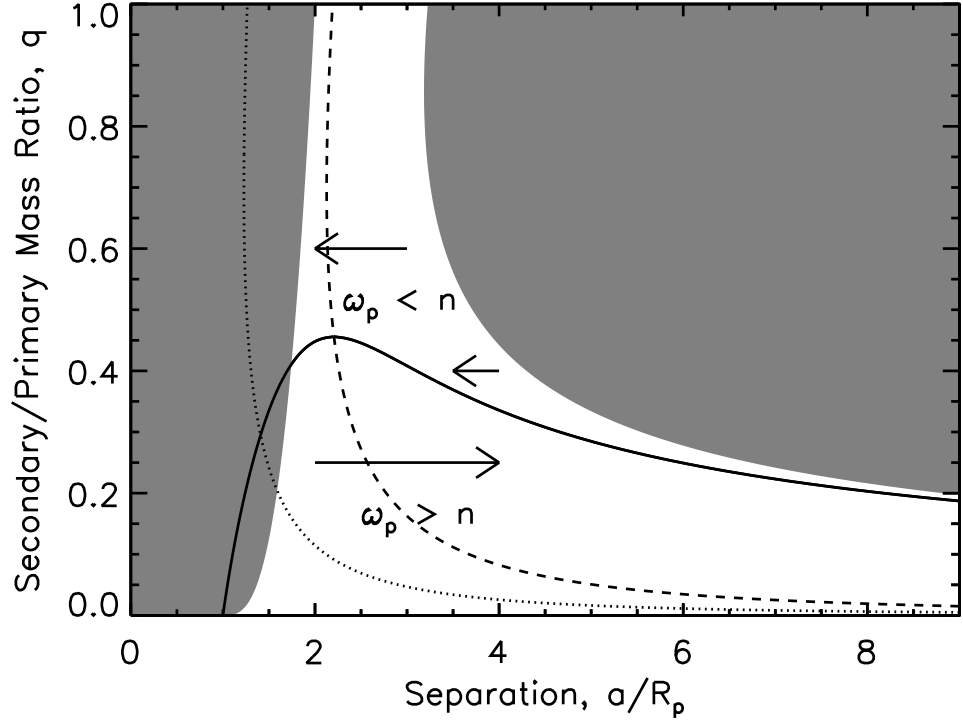}
\caption[]{Directions of tidal evolution for systems with $J/J^{\prime}=0.4$.  For binaries dominated 
by the spin angular momentum of the primary mass, systems under the fully synchronous orbit curve will 
evolve outward as angular momentum is transferred from the rapid spin of the primary to the orbit 
($\omega_{\rm p}>n$).  Systems above the fully synchronous orbit curve evolve inward as angular momentum 
is transferred from the orbit to the spins of the components ($\omega_{\rm p}<n$) until either the outer 
synchronous orbit is reached or the orbit collapses to contact.  The shaded regions indicate separations 
that are inaccessible to the binary components due to the contact limit and angular momentum limit for 
$J/J^{\prime}=0.4$.  The dashed line indicates the synchronous stability limit, while the dotted line 
indicates where $E=0$ for fully synchronous orbits (see Section~\ref{sec:stablimit}).}
\label{fig:synchdir}
\end{center}
\end{figure}

Figure~\ref{fig:synch} is a hybrid of the ``universal diagram'' for tidal evolution of~\citet{coun73} and 
Fig.~1 of~\citet{weid89} in that it combines the ability to directly read off the fully synchronous orbits 
for specific values of angular momentum of Counselman with the mass ratio dependence of Weidenschilling et al.  
Furthermore, our figure accounts for the spin of the secondary, which Counselman does not, and can be modified 
to include tidal evolution timescales as done by Weidenschilling et al., which we will do in 
Section~\ref{sec:material}.  Understanding evolution through our version of the universal diagram is aided by 
Fig.~\ref{fig:synchdir}, which illustrates the general trends of tidal evolution when the spins of the 
components have the same sense of rotation as the motion in the mutual orbit, \textit{i.e.}, all motion is
prograde.  If the configuration of the system falls between the two solutions to the fully synchronous orbit 
equation for the $q$ of the system, meaning it visually appears ``below'' the fully synchronous orbit curve in 
$\left(q, a/R_{\rm p}\right)$-space like the lower arrow in Fig.~\ref{fig:synchdir}, the system will evolve 
outward as spin angular momentum is transferred to the mutual orbit via the tidal interaction 
($\omega_{\rm p}>n$, $\omega_{\rm s}\sim n$).  If the system sits ``above'' its fully synchronous orbit curve
in the position of one of the upper arrows in  Fig.~\ref{fig:synchdir}, due to having a large secondary or a 
wide separation, the orbital angular momentum makes up a larger fraction of the total angular momentum such 
that the angular momentum available in the spins of the components requires that $\omega_{\rm p}<n$ (for 
$\omega_{\rm s}\sim n$).  The system must then evolve inward as angular momentum from the orbit is transferred 
to the spins of the components.  For systems with constant angular momentum, the cases of inward evolution 
require a binary formation mechanism that initially produces well-separated components with rotation rates 
slower than the mean motion because tidal evolution cannot have evolved a system outward to these 
configurations above the fully synchronous orbit curve.  Note that binaries with equal-mass $\left(q=1\right)$ 
components in Fig.~\ref{fig:synch} only have fully synchronous end states if $J/J^{\prime}>0.44$.  For 
$J/J^{\prime}<0.44$, cases can exist where $q$ lies entirely above the fully synchronous orbit curve (the 
uppermost arrow in Fig.~\ref{fig:synchdir}) meaning the fully synchronous orbit equation has no solution for 
that value of $q$.  The resulting inward tidal evolution may be important in the context of forming contact 
binary asteroids~\citep{tayl09dda} and will be addressed in the future.


\section{Stability and $E=0$ limits of fully synchronous orbits}
\label{sec:stablimit}

The fully synchronous orbit solutions to Eq.~\ref{eq:synch} are equivalently thought of as contours of 
constant $J/J^{\prime}$ in ($q$, $a/R_{\rm p}$)--space satisfying:

\begin{equation}
\frac{J}{J^{\prime}}~=~\frac{1}{\left(1+\frac{\rho_{\rm p}}{\rho_{\rm s}}q\right)^{1/6}}\,\frac{q}{\left(1+q\right)^2}\,\left(\frac{a}{R_{\rm p}}\right)^{1/2}+\alpha_{\rm p}\,\frac{1+\frac{\alpha_{\rm s}}{\alpha_{\rm p}}\left(\frac{\rho_{\rm p}}{\rho_{\rm s}}\right)^{2/3}q^{5/3}}{\left(1+\frac{\rho_{\rm p}}{\rho_{\rm s}}q\right)^{1/6}\left(1+q\right)}\,\left(\frac{a}{R_{\rm p}}\right)^{-3/2}.
\label{eq:jcontour}
\end{equation}

\noindent
The maxima of the set of contour curves for a continuous range of $J/J^{\prime}$ are traced out by taking
the derivative of Eq.~\ref{eq:jcontour} with respect to $a/R_{\rm p}$ giving:

\begin{equation}
\frac{a_{\rm stab}}{R_{\rm p}}~=~\left[3\alpha_{\rm p}\,\frac{1+q}{q}\left(1+\frac{\alpha_{\rm s}}{\alpha_{\rm p}}\left(\frac{\rho_{\rm p}}{\rho_{\rm s}}\right)^{2/3}q^{5/3}\right)\right]^{1/2}.
\label{eq:stab}
\end{equation}

\noindent
For spherical components with similar uniform densities: 

\begin{equation}
\frac{a_{\rm stab}}{R_{\rm p}}~=~\left[\frac{6}{5}\,\frac{1+q}{q}\,\left(1+q^{5/3}\right)\right]^{1/2},
\end{equation}

\noindent
which is the dashed curve in Fig.~\ref{fig:synch} that splits the synchronous orbit curves into inner
and outer solutions when there are two real solutions to Eq.~\ref{eq:synchsphere}.  \citet{harr82} showed 
that Eq.~\ref{eq:stab} is also a stability limit against perturbations to the system such that the outer 
synchronous orbit in the two-sphere case that satisfies:

\begin{equation}
\frac{a}{R_{\rm p}}~>~\left[\frac{6}{5}\,\frac{1+q}{q}\,\left(1+q^{5/3}\right)\right]^{1/2}
\end{equation}

\noindent
is stable, while the inner synchronous orbit is not, meaning that the binary system will tend to evolve away
from the inner synchronous orbit given the opportunity.  If perturbed outward from the inner synchronous 
orbit, the components will tidally evolve outward to the stable outer synchronous orbit (the lower arrow in 
Fig.~\ref{fig:synchdir}); if perturbed inward, the components will fall to contact, though for most 
$J/J^{\prime}$, the inner synchronous orbit is already within the contact limit.  The arrows in 
Fig.~\ref{fig:synchdir} also show that perturbation of the system from the outer synchronous orbit will allow 
the system to return to the outer synchronous orbit, indicating its stability against tidal perturbation.  It 
is also clear from Figs.~\ref{fig:synch} and~\ref{fig:synchdir}, that the stability limit is always beyond 
the contact limit such that whenever a particle is lofted from the surface of the primary or when a parent 
body fissions, the components are initially in a tidally unstable configuration that may lead to 
re-impact~\citep{sche09}.  

For bodies with $\alpha_{\rm p, s}\ne2/5$, one may use the parameter $\lambda_{\rm p, s}=5\alpha_{\rm p, s}/2$ 
introduced by~\citet{desc08} to find that the stable outer synchronous orbit is described by:

\begin{equation}
\frac{a}{R_{\rm p}}~>~\left[\frac{6}{5}\,\frac{1+q}{q}\,\left(\lambda_{\rm p}+\lambda_{\rm s}q^{5/3}\right)\right]^{1/2},
\label{eq:marchis}
\end{equation}

\noindent
correcting both the formula and direction of the inequality given by~\citet{marc08ecc,marc08circ}.  We note, 
though, that if one (or both) of the components of the binary system is nonspherical, Eq.~\ref{eq:marchis} is not 
completely accurate because the formulation of the fully synchronous orbit curves in Eq.~\ref{eq:synch}, and thus 
the stability limit, would be complicated by the mutual gravitational potential between nonspherical 
bodies~\citep{sche09} and the synchronous rotation rate required for relative equilibrium~\citep[see][]{sche06,sche07}.  
Therefore, Eq.~\ref{eq:marchis} is exact only in the two-sphere case with $\alpha_{\rm p, s}\ne2/5$ due to density 
inhomogeneity within the individual components.

The stability limit in Eq.~\ref{eq:stab} gives the minimum separation of a stable, fully synchronous 
tidal end state for a given mass ratio $q$.  The corresponding minimum angular momentum required for a system to 
have a stable tidal end state results from substituting Eq.~\ref{eq:stab} into Eq.~\ref{eq:jcontour} and, by 
assuming similar component densities and $\alpha_{\rm p, s}=2/5$, can be written as:

\begin{equation}
J/J^{\prime}_{\rm crit}~=~\left(\frac{512}{135}\right)^{1/4}\frac{q^{3/4}}{\left(1+q\right)^{23/12}}\left(1+q^{5/3}\right)^{1/4}.
\label{eq:jmin} 
\end{equation}

\noindent
This critical value for the angular momentum is equivalent to that derived by~\citet{hut80} in the context of 
binary star systems.  Satisfying the stability condition is a necessary, but not sufficient, condition for a stable 
end to tidal evolution.  Systems that do not satisfy the stability condition have not reached a stable tidal end 
state, while systems that do satisfy the stability condition may or may not continue their evolution depending on 
the angular momentum of the system.  At the minimum angular momentum of Eq.~\ref{eq:jmin}, the synchronous orbit 
equation (Eq.~\ref{eq:synchsphere}) has one degenerate solution, which is simply the stability limit, and tidal 
evolution may cease.  If the angular momentum of the system exceeds $J/J^{\prime}_{\rm crit}$, the synchronous 
orbit equation has two solutions, and the system will continue to evolve to the outer synchronous orbit.  Of 
course, having less than $J/J^{\prime}_{\rm crit}$ will cause the system to collapse as the synchronous orbit 
equation has no solution regardless of whether or not the stability condition is satisfied.  This argument is used 
by~\citet{levr09} to posit that nearly all transiting extrasolar planets, in the absence of a parking mechanism, 
will inevitably collide with their host stars because these planetary systems lack sufficient angular momentum to 
support stable tidal end states for their components.

Upon full synchronization $\left(\omega_{\rm p}=\omega_{\rm s}=n\right)$, the total energy of the system 
(Eq.~\ref{eq:Etot}) may be positive or negative with the $E=0$ limit occurring at:

\begin{equation}
\frac{a_{\rm E}}{R_{\rm p}}~=~\left[\alpha_{\rm p}\,\frac{1+q}{q}\left(1+\frac{\alpha_{\rm s}}{\alpha_{\rm p}}\left(\frac{\rho_{\rm p}}{\rho_{\rm s}}\right)^{2/3}q^{5/3}\right)\right]^{1/2},
\label{eq:energy}
\end{equation}

\noindent
where spherical shapes have been assumed in determining the mutual gravitational potential.  Fully synchronous 
orbits with semimajor axes $a_{\rm sync}>a_{\rm E}$ have $E<0$ because the combination of translational and 
potential energy $\left(\propto a^{-1}\right)$ falls off more slowly with increasing $a$ than the rotational 
energy of the components $\left(\propto a^{-3}\right)$ when synchronized.  Fully synchronous orbits with 
$a_{\rm sync}<a_{\rm E}$ have $E>0$ and could evolve to escape given sufficient perturbation from relative 
equilibrium~\citep{bell08}.  Note that in the two-sphere case $a_{\rm E}=a_{\rm stab}/\sqrt{3}$ such that all 
stable, fully synchronous orbits have $E<0$, but the converse, having $E<0$, does not necessarily guarantee 
stability.  For systems with $J/J^{\prime}>0.334$ in Fig.~\ref{fig:synch}, the realistic fully synchronous 
orbit solutions, \textit{i.e.}, those beyond the contact limit, have $E<0$.  Systems with less angular momentum 
can have unstable inner synchronous orbits for which $E>0$.  

As an interesting side note, if one inserts $a_{\rm E}$ into $L$ and $S$ in Eq.~\ref{eq:Lkepler} and Eq.~\ref{eq:S} 
for a fully synchronous orbit, one finds that $L/S=1$ or the angular momentum of the system revolving about its 
center of mass is equal to the rotational angular momentum of the components~\citep{sche09}.  Similarly, inserting 
$a_{\rm stab}$ finds $L/S=3$ or the angular momentum of the system's revolution is three times the rotational 
angular momentum~\citep{hut80,sche09}.  Thus, a stable fully synchronous tidal end state must satisfy $L/S \ge 3$ 
or, equivalently, the orbital angular momentum must account for more than $3/4$ of the total angular momentum of 
the system, which is observed in nearly equal-mass binary systems (90) Antiope~\citep{mich04,desc07} and (69230) 
Hermes~\citep{marg06iau}.


\section{Material properties of binary asteroids}
\label{sec:material}

During the tidal evolution process, angular momentum is transferred between the spins of the components 
$\omega_{\rm p}$ and $\omega_{\rm s}$ and the mutual orbit about the center of mass of the system.  For 
a prograde system, angular momentum conservation~\citep[\textit{e.g.},][]{murr99,tayl10cmda} requires 
that the change with time of the orbital separation of the components follows:

\begin{eqnarray}
\frac{\dot{a}}{R_{\rm p}} & = & \frac{8\sqrt{3}}{19}\,\frac{\pi^{3/2}G^{3/2}\rho_{\rm p}^{5/2}R_{\rm p}^2}{\mu_{\rm p}Q_{\rm p}}\,q\left(1+q\right)^{1/2}\left(\frac{a}{R_{\rm p}}\right)^{-11/2}\nonumber\\
 & & \times\,\left[{\rm sign}\left(\omega_{\rm p}-n\right)+\frac{R_{\rm s}}{R_{\rm p}}\,\frac{\mu_{\rm p}Q_{\rm p}}{\mu_{\rm s}Q_{\rm s}}\,{\rm sign}\left(\omega_{\rm s}-n\right)\right].
\label{eq:adot}
\end{eqnarray}

\noindent
The sign of the quantity $\omega-n$ for each component determines whether the separation increases or 
decreases due to tides raised on that component; if $\omega>n$ ($\omega<n$), tides slow (accelerate) the 
rotation of the component and increase (decrease) the orbital separation.  The contribution from tides 
raised on the secondary are a factor of the size ratio weaker than from tides raised on the primary and, 
furthermore, the smaller secondary will tend to synchronize to the mean motion $n$ before the primary 
causing the tidal torque on the secondary, as well as its contribution to the change in orbital separation, 
to vanish earlier.  The overall strength of the tidal mechanism is inversely related to the product 
$\mu Q$, which combines the rigidity or shear modulus $\mu$ of the material\footnote{In deriving 
Eq.~\ref{eq:adot}, it is assumed that for both components the rigidity $\mu$ dominates the stress due 
to self-gravity $\rho g R\,\sim\,G\rho^{2}R^{2}$, where $g$ is the surface gravity, which is reasonable 
for bodies under 200 km in radius~\citep{weid89} such that the potential Love number $k_{2} \propto 
\mu^{-1}$.} and the tidal dissipation function $Q$ that relates to the lag angle between the axis of 
symmetry of the tidal bulge and the tide-raising body~\citep[see][for detailed discussions]{gold63,efro09}.  
The larger the quantity $\mu Q$, the more resistant the material is to orbit evolution due to the tidal 
mechanism.  Here, for completeness, we have allowed for the primary and secondary to have their own 
$\mu Q$ values. 

\subsection{Caveats}
\label{sec:caveats}

In estimating the material properties of binary asteroids, we assume that tides are the dominant method
of dynamical evolution in these systems, specifically in the evolution of the separation between the 
components.  It has been argued that the binary YORP effect~\citep[BYORP;][]{cuk05,cuk10,mcma10cmda}, 
where a synchronous secondary acts to asymmetrically re-radiate sunlight with respect to its orbital velocity 
so that the orbit is expanded or contracted, can act on timescales faster than tidal 
evolution~\citep{cuk05,gold09,mcma10icarus}.  This mechanism is similar to the YORP effect~\citep{rubi00}, 
an asymmetric re-radiation of sunlight with respect to rotational velocity.  However, unlike the YORP 
effect~\citep{tayl07,lowr07,kaas07}, BYORP has yet to be proven observationally.  Furthermore, BYORP 
requires a synchronous secondary and thus cannot be the dominant active evolution mechanism for systems 
with asynchronous secondaries.  Also of interest, specifically among small binary asteroids with primaries less 
than 10~km in diameter, is the idea of mass lofting or ``tidal saltation''~\citep{harr09}, where particles at 
the equator of a rapidly spinning primary become weightless due to the gravitational presence of the secondary 
passing overhead, briefly enter orbit, and transfer angular momentum to the orbit of the secondary before 
falling back to the surface of the primary.  This method has also been argued to expand the mutual orbit more 
quickly than tidal evolution~\citep{fahn09}.  It may be a combination of effects, also including close planetary 
flybys~\citep{fari92,bott96a,bott96b}, that evolve the separation of near-Earth binary components, but 
tides are the only mechanism one can say at this time must act on all systems.  For main-belt binaries 
with 100-km-scale primaries, tides should be the dominant mechanism because close planetary encounters 
are not feasible in the main belt, the thermal YORP and BYORP effects are weakened by the increased 
heliocentric distance to the main belt (and the increased sizes of the bodies involved), and the primaries 
do not rotate rapidly enough nor are the secondaries close enough to produce mass lofting.

\subsection{Estimation of $\mu Q$}
\label{sec:muq}

Integration of Eq.~\ref{eq:adot} provides, symbolically, the evolution of the orbital separation as a 
function of time in terms of the product $\mu_{\rm p}Q_{\rm p}$ and the time over which tidal evolution 
has taken place $\Delta t$.  Both $\mu_{\rm p}Q_{\rm p}$ and $\Delta t$ are inherent unknowns in any binary 
system, but their ratio is fully determined by the current separation of the binary system $a/R_{\rm p}$, 
the physical properties of the components, and the assumption of an initial separation of the components after 
the formation of the binary.  The ratio is very weakly dependent on the initial separation (see 
Section~\ref{sec:other}); we choose 2$R_{\rm p}$ because it is both the contact limit for equal-size components 
and a reasonable initial separation for binaries formed via spin-up~\citep{wals08yorp}.  Thus, the evolution 
of the orbital separation gives:

\begin{equation}
\frac{\mu_{\rm p}Q_{\rm p}}{\Delta t}~=~\frac{52\sqrt{3}}{19}\,\pi^{3/2}G^{3/2}\rho_{\rm p}^{5/2}R_{\rm p}^{2}\,q\left(1+q\right)^{1/2}\left[\left(\frac{a}{R_{\rm p}}\right)^{13/2}-2^{13/2}\right]^{-1}\,\left(1+\frac{R_{\rm s}}{R_{\rm p}}\,\frac{\mu_{\rm p}Q_{\rm p}}{\mu_{\rm s}Q_{\rm s}}\,\frac{\tau_{\rm s}}{\Delta t}\right)
\label{eq:muq}
\end{equation}

\noindent
for a system where tides on both components are causing the mutual orbit to expand with $\tau_{\rm s}$ 
representing the length of time the secondary contributes to the tidal evolution, which is less than or 
equal to the age of the binary system $\Delta t$.  Ignoring the rightmost term in parentheses for the 
moment, to estimate either $\mu_{\rm p}Q_{\rm p}$ or $\Delta t$, the other must be assumed, where varying 
one by, say, an order of magnitude varies the other parameter by the same amount.  With a judicious 
assumption of the age of the binary, one can estimate $\mu_{\rm p}Q_{\rm p}$ for the primary or at least 
place rough bounds on its value~\citep{marg02s,marg03}; similarly, by assuming a value of $\mu_{\rm p}Q_{\rm p}$, 
one can estimate the age of the binary~\citep{wals06,marc08ecc,marc08circ,gold09}.  A related analysis was 
done for the martian moon Phobos by~\citet{yode82}, who estimated its $\mu Q$ is at least of order 10$^{11}$ 
N\,m$^{-2}$ based on the eccentricity evolution of Phobos within the $\Delta t$ inferred from the minimum 
age of its surface.

If one ignores the contribution of the secondary, equivalent to setting $\tau_{\rm s}$ to zero, Eq.~\ref{eq:muq} 
represents a lower bound on $\mu_{\rm p}Q_{\rm p}$ for a given timescale $\Delta t$.  Allowing the secondary to 
contribute for the entire age of the binary by letting $\tau_{\rm s}\rightarrow\Delta t$ gives an upper 
bound\footnote{We do not use the terms ``lower bound'' and ``upper bound'' in a rigorous mathematical sense here
because factors such as uncertainties in the physical properties used in Eq.~\ref{eq:muq} can affect the values 
of these bounds by factors of order unity.} on $\mu_{\rm p}Q_{\rm p}$, which is a factor of 
$1+\left(R_{\rm s}/R_{\rm p}\right)\left(\mu_{\rm p}Q_{\rm p}/\mu_{\rm s}Q_{\rm s}\right)$ greater than the 
lower bound found using only the primary.  If $\Delta t$ is correct, the actual $\mu_{\rm p}Q_{\rm p}$ should 
lie between the two bounds as the secondary tends to be synchronously rotating such that $\tau_{\rm s}<\Delta t$ 
rather than contributing to the mutual orbit expansion over the entire age of the system.  Ideally, one would 
know when the secondary became synchronized, evolve the system using both tides until $\tau_{\rm s}$, then 
include only the tide on the primary for the remainder of $\Delta t$.  Our ignorance of if/when the secondary 
became synchronous cannot change the estimate of $\mu_{\rm p}Q_{\rm p}$ by more than a factor of 
$1+\left(R_{\rm s}/R_{\rm p}\right)\left(\mu_{\rm p}Q_{\rm p}/\mu_{\rm s}Q_{\rm s}\right)$, which is nominally 
less than two for components with equivalent material properties, far less critical than the choice of $\Delta t$, 
the age of the binary, but similar to the effect of measurement errors on the physical parameters in 
Eq.~\ref{eq:muq} (see Section~\ref{sec:other}).

In examining orbit expansion by tides, one cannot directly determine whether the primary and secondary have 
different $\mu Q$ values, though having a stronger or weaker secondary can change the estimate of 
$\mu_{\rm p}Q_{\rm p}$.  Assuming equivalent $\mu Q$ values for the components, $\mu_{\rm p}Q_{\rm p}$ cannot be 
affected by more than a factor of two by accounting for the secondary; often the factor will be much less due to 
the size ratio of the components and the fact that synchronous secondaries require $\tau_{\rm s}<\Delta t$.  If 
the secondary is mechanically stronger than the primary such that $\mu_{\rm s}Q_{\rm s}>\mu_{\rm p}Q_{\rm p}$, 
the effect on $\mu_{\rm p}Q_{\rm p}$ compared to ignoring the secondary in Eq.~\ref{eq:muq} must still be less 
than a factor of two because 
$\left(R_{\rm s}/R_{\rm p}\right)\left(\mu_{\rm p}Q_{\rm p}/\mu_{\rm s}Q_{\rm s}\right)\left(\tau_{\rm s}/\Delta t\right)<1$.  
Thus, having a stronger secondary than the primary serves to reduce the upper bound on $\mu_{\rm p}Q_{\rm p}$ for a 
specific age $\Delta t$, but the actual material strength of the secondary is not constrained.  On the other hand, 
if the secondary is mechanically weaker than the primary such that $\mu_{\rm s}Q_{\rm s}<\mu_{\rm p}Q_{\rm p}$, 
depending on the size and $\mu Q$ ratios of the components and the age of the system $\Delta t$, the effect on 
$\mu_{\rm p}Q_{\rm p}$ compared to ignoring the secondary could conceivably exceed a factor of two, but is 
unlikely to differ by more than an order of magnitude unless the system is very young and the secondary is orders 
of magnitude weaker than the primary.\footnote{The time it takes for the secondary to synchronize $\tau_{\rm s}$ 
shortens compared to $\Delta t$ as $\mu_{\rm s}Q_{\rm s}$ is made weaker so that $\tau_{\rm s}/\Delta t$ tends to 
cancel out the effect of $\mu_{\rm p}Q_{\rm p}/\mu_{\rm s}Q_{\rm s}$ growing.  Only if $\tau_{\rm s}$ and 
$\Delta t$ are comparable due to the youth of the system will the disparity between the mechanical strength of 
the primary and secondary have a strong effect on the calculation at hand (if it can overcome the size ratio 
factor as well).}

For a synchronous secondary, the eccentricity can increase or decrease depending on how $\mu_{\rm s}Q_{\rm s}$ 
compares to $\mu_{\rm p}Q_{\rm p}$~\citep{gold63,gold66} with the prediction that, for equivalent material properties, 
systems with monolithic components and $q \le 0.31$ will have the eccentricity excited via tides~\citep{harr82,weid89}.  
Given we observe the vast majority of binaries, including those with much smaller mass ratios than 0.31, have 
circular mutual orbits, to prevent eccentricity excitation the secondaries must be mechanically weaker than the 
primaries~\citep{marg03} by roughly an order of magnitude to lower the limiting mass ratio from 0.31 to below the 
observed mass ratios in the binary population, \textit{e.g.}, by a factor of 30 for the smallest mass-ratio system 
we will discuss, (702) Alauda.  However, \citet{gold09} suggest that binary systems with gravitational-aggregate 
components, rather than monolithic ones, do not require much weaker secondaries to have circular mutual orbits as the 
eccentricity will damp for all mass ratios if the components are aggregates of similar material.  So again, it 
is difficult to constrain the separate values of $\mu Q$ for the components, but our ignorance should not strongly 
affect our results for particular systems unless the extreme case of a very young binary with components having very 
disparate material properties exists.  Therefore, when estimating the material properties of a binary system via 
Eq.~\ref{eq:muq}, we may ignore the possibility of the components having different material properties to obtain an 
order-of-magnitude result.\footnote{By ignoring any possible difference in $\mu Q$ of the components, Eq.~\ref{eq:muq} 
gives the value of $\mu Q$ for the system, as in it applies to both components rather than the primary only.}  In 
the following sections we will also ignore the contribution of the secondary in Eq.~\ref{eq:muq} to find a lower 
bound for $\mu Q$ of the system with a specific age $\Delta t$.  An upper bound on $\mu Q$ from accounting for the 
secondary is nominally less than a factor of two larger if the material properties of the components are similar.

\subsection{Large main-belt binaries}
\label{sec:mba}

First we consider binary systems in the main asteroid belt and among the Jupiter Trojans with 100-km-scale 
primary components that have been characterized by direct adaptive-optics imaging and lightcurve photometry.  
To be included in Table~\ref{tab:mba}, orbital properties must be known and some estimate of the sizes must be 
available.  Most of the binary systems discovered via imaging of large main-belt asteroids and Jupiter Trojans 
have secondaries roughly one-tenth the size of their primaries ($q\sim0.001$) or smaller.  These secondaries 
are likely the result of a sub-catastrophic impact on the parent 
body (\textit{i.e.}, SMATS, smashed target satellites, as described by~\citet{durd04}), which is supported by 
the angular momentum budget of the systems.  With the exception of (90) Antiope and (617) Patroclus, which have 
nearly equal-size components, the 100-km scale main-belt binaries have $J/J'$ values of roughly 0.2, well below 
the $J/J^{\prime}\sim0.4$ regime characteristic of binaries formed via a spin-up mechanism.  Among the $q<0.1$ 
binaries, the average primary component spins with a period of roughly 6 h, twice the rotational breakup 
period, yet accounts for 97\% of the angular momentum of the system due to the large mass disparity between the 
components.  Nearly equal-mass binary (90) Antiope~\citep{merl00iauc,mich04,desc07} has $J/J^{\prime}\sim0.5$ 
that is more similar to a binary formed through spin-up than through a sub-catastrophic collision.  However, the 
sheer size of the components, each more than 80 km in diameter, and its location in the main belt make it difficult 
for the YORP effect or close planetary encounters to explain how the (90) Antiope system originally formed.  
Nearly equal-mass binary (617) Patroclus~\citep{merl01,marc06,muel10} has $J/J^{\prime}\sim0.8$, far larger than 
the other large main-belt binaries we consider, but near the upper limit for giant impacts of 
$J/J^{\prime}<0.8$~\citep{canu05}.  Such a high angular momentum content is more similar to many of the 
Kuiper-belt binaries that may have formed via $n$-body capture~\citep[see][for a review]{noll08} rather than 
spin-up or collisions.


\begin{table}[!p]\normalsize
\begin{center}
\begin{tabular}{rlccccccc}
$ $ & Name & $R_{\rm p}$ (km) & $R_{\rm s}/R_{\rm p}$ & $q$ & $\rho$ (g~cm$^{-3}$) & $a/R_{\rm p}$ & $J/J^{\prime}$ & $\mu Q$ (N\,m$^{-2}$) \\
\\
 22 & Kalliope        &  85 & 0.213 & 0.009664 & 2.5 & 12.5 & 0.22 & 3.3\,$\times$\,10$^{12}$\\
 45 & Eugenia         &  98 & 0.036 & 0.000047 & 1.1 & 12.1 & 0.19 & 3.4\,$\times$\,10$^{9}$\\
 87 & Sylvia          & 128 & 0.063 & 0.000250 & 1.5 & 10.6 & 0.20 & 1.6\,$\times$\,10$^{11}$\\
 90 & Antiope$^{a}$   &  43 & 0.955 & 0.871    & 1.3 &  3.9 & 0.49 & $<$\,4.0\,$\times$\,10$^{16}$\\
107 & Camilla         & 103 & 0.050 & 0.000125 & 1.9 & 12.0 & 0.18 & 4.1\,$\times$\,10$^{10}$\\
121 & Hermione        & 103 & 0.066 & 0.000287 & 1.1 &  7.5 & 0.22 & 5.1\,$\times$\,10$^{11}$\\
130 & Elektra         &  90 & 0.026 & 0.000018 & 3.0 & 14.0 & 0.14 & 5.1\,$\times$\,10$^{9}$\\
283 & Emma            &  73 & 0.079 & 0.000493 & 0.8 &  8.2 & 0.20 & 1.1\,$\times$\,10$^{11}$\\
617 & Patroclus$^{b}$ &  51 & 0.920 & 0.779    & 1.3 & 13.5 & 0.83 & 1.5\,$\times$\,10$^{13}$\\
702 & Alauda          &  97 & 0.018 & 0.000006 & 1.6 & 12.6 & 0.13 & 7.9\,$\times$\,10$^{8}$\\
762 & Pulcova         &  67 & 0.160 & 0.004096 & 1.9 & 12.2 & 0.17 & 5.1\,$\times$\,10$^{11}$\\
\end{tabular}
\end{center}
\caption[]{Physical properties of main-belt and Jupiter-Trojan binary asteroids with primary radii of order 
100 km.  $R_{\rm p}$ is the radius of the primary; $R_{\rm s}/R_{\rm p}$ is the size ratio of the components; 
$q$ is the mass ratio, taken to be the size ratio cubed by assuming components of equal density $\rho$; 
$a/R_{\rm p}$ is the semimajor axis of the mutual orbit; $J/J^{\prime}$ is the normalized angular momentum 
assuming spherical bodies and a synchronous secondary using Eqs.~\ref{eq:Lkepler} and~\ref{eq:S}, for which 
an average value of roughly 0.2 for $q < 0.1$ binaries indicates a sub-catastrophic collisional origin rather 
than formation through spin-up.  For the triple systems (45) Eugenia and (87) Sylvia, only properties of the 
larger outer satellite are considered.  Values of $\mu Q$ correspond to tides raised on the primary acting over 
the 4.5 Gyr age of the Solar System.  Accounting for tides raised on the secondary would increase $\mu Q$ by no 
more than a factor of two.  Data are taken from the Ondrejov Asteroid Photometry Project binary asteroid 
parameters table (http://www.asu.cas.cz/$\sim$asteroid/binastdata.htm, 2010 April 8 release) with the exception 
of (702) Alauda whose parameters are provided by~\citet{rojo11}.\\$^{a}$ (90) Antiope is known to be in a fully 
synchronous state where tidal evolution has ceased.  Because of its nearly equal-size components in such close 
proximity, its tidal evolution occurred over much less than the age of the Solar System.\\ $^{b}$ (617) 
Patroclus is believed to be in a fully synchronous state and must have formed at a larger initial separation 
than the value of $2R_{\rm p}$ assumed here.}
\label{tab:mba}
\end{table}

\begin{figure}[!p]
\begin{center}
\includegraphics[angle=90., scale=0.60]{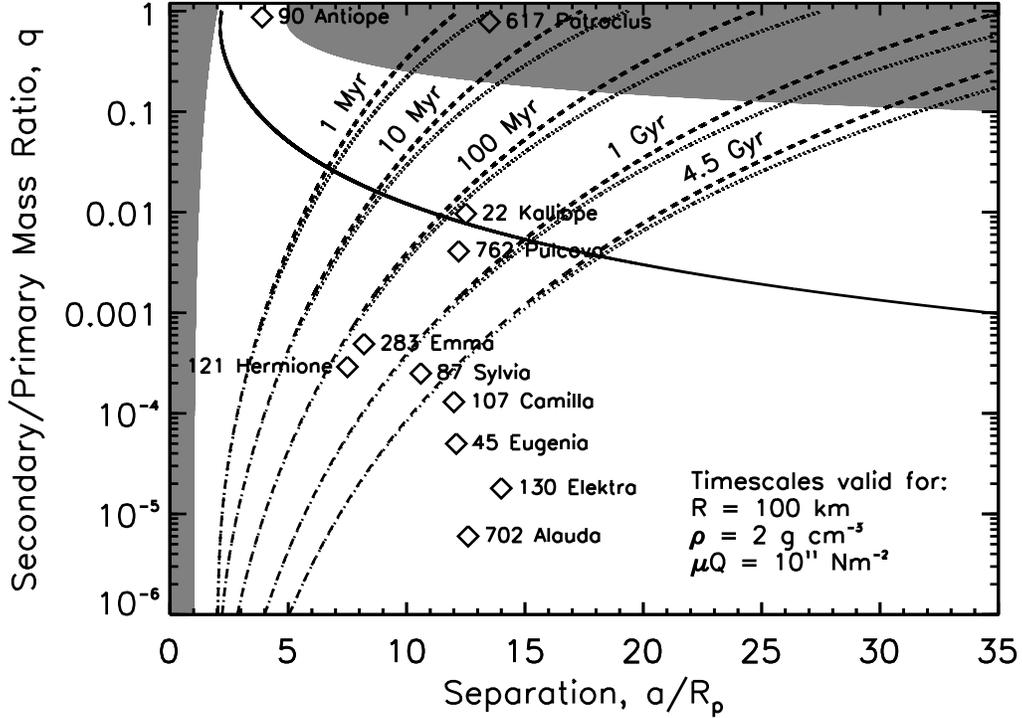}
\caption[]{Mass ratio $q$ and primary-secondary separation $a/R_{\rm p}$ for 100-km-scale main-belt 
and Jupiter-Trojan binaries along with the angular momentum limit for $J/J^{\prime}=0.5$.  The solid curve 
is the stability limit.  Tidal evolution timescales are plotted assuming $\rho_{\rm p, s}=2$ g~cm$^{-3}$, 
$R_{\rm p}=100$ km, and $\mu_{\rm p}Q_{\rm p}=\mu_{\rm s}Q_{\rm s}=10^{11}$ N\,m$^{-2}$ and allowing tides 
raised on the primary only (dashed) or both components (dotted) to contribute over the entire evolution 
from an initial separation of $2R_{\rm p}$.  With these parameters, (90) Antiope evolves far more rapidly 
than the other main-belt binaries.}
\label{fig:mba}
\end{center}
\end{figure}

The $q<0.1$ binaries in the main belt are in the midst of a lengthy tidal evolution.  As shown in 
Fig.~\ref{fig:mba}, of the $q<0.1$ binaries, only (22) Kalliope has reached the synchronous stability limit, 
but even (22) Kalliope is very far from reaching the outer synchronous orbit that nearly coincides with the 
angular momentum limit for these systems.  The inner synchronous orbit for $J/J^{\prime}=0.2$ lies above the 
contact limit, so these binaries must have begun their tidal evolution from an initial separation of roughly 
2$R_{\rm p}$ or more to have evolved outward due to tides.  Nearly equal-mass binaries (90) 
Antiope~\citep{mich04,desc07} and (617) Patroclus~\citep{marc06,muel10} are both believed to be in their fully 
synchronous tidal end states where the rotational periods of the components equal the period of the mutual orbit.  
Of the large main-belt binaries listed in Table~\ref{tab:mba}, only (130) Elektra and (283) Emma have mutual 
orbits that are not roughly circular, each having an eccentricity $e\sim 0.1$ likely caused by tidal 
excitation~\citep{marc08ecc}.


Assuming the value of $\mu Q$ of $10^{11}$ N\,m$^{-2}$ for Phobos determined by~\citet{yode82} is applicable 
to the components of main-belt binaries, timescales for tidal evolution are plotted in Fig.~\ref{fig:mba} and
illustrate the interplay between material strength and system age.  Systems that plot to the right of the 
4.5 Gyr curve, the age of the Solar System, must either have smaller $\mu Q$ values than Phobos or have evolved
very little in separation since formation. Systems that plot to the left of the 4.5 Gyr curve are either 
younger than 4.5 Gyr or must have $\mu Q$ values larger than $10^{11}$ N\,m$^{-2}$ if they are actually 4.5 Gyr 
old.  Solid rock has a rigidity $\mu$ of $10^{10}$ N\,m$^{-2}$ or greater~\citep{gold66,dzie81} and, assuming 
$Q\gg1$, the product $\mu Q$ for solid rock would be of order $10^{12}$ N\,m$^{-2}$ or greater;~\citet{burn73} 
adopt a $\mu Q$ value of $3 \times 10^{13}$ N\,m$^{-2}$ as an extreme case for solid, non-porous rock in their 
analysis of damping to principal axis rotation.  Furthermore,~\citet{harr94} argues that $\mu Q$ should fall 
within a factor of 100 of $5 \times 10^{11}$ N\,m$^{-2}$ based on analyses of Phobos and Comet 
Halley~\citep{peal89}.  In summary, a monolithic rock may have $\mu Q$ of order $10^{12}$ N\,m$^{-2}$ or greater 
while a fractured rock would be expected to have a lower $\mu Q$ similar to that of Phobos with $\mu Q$ 
continuing to decrease with increased fracturing~\citep{he94}.

If instead of assuming a value of $\mu Q$ for the binary systems, we set $\Delta t$ to 4.5 Gyr, the age of the 
Solar System and the maximum amount of time these systems could have been tidally evolving, we can place an 
upper bound\footnote{When considering a specific value of $\Delta t$, Eq.~\ref{eq:muq} provides a lower bound 
when ignoring the secondary and an upper bound when accounting for the secondary over the entire age $\Delta t$.  
Here, we are considering all possible values of $\Delta t$ such that using the maximum possible age will give 
the maximum $\mu Q$ value to within a factor of two using Eq.~\ref{eq:muq} and accounting only for tides on the 
primary component.} on $\mu Q$ with Eq.~\ref{eq:muq}.  Ignoring (90) Antiope and (617) Patroclus, which can 
tidally evolve on timescales much more rapid than the age of the Solar System due to having nearly equal-size 
components, the median $\mu Q$ from Table~\ref{tab:mba} for main-belt binaries with 100-km-scale primaries is 
$1 \times 10^{11}$ N\,m$^{-2}$, the same value estimated by~\citet{yode82} for Phobos and reasonable for 
fractured rock.  Individual values of $\mu Q$ are roughly consistent with the prediction of~\citet{harr94} and 
range over more than three orders of magnitude from $7.9 \times 10^{8}$ N\,m$^{-2}$ for (702) Alauda to 
$3.3 \times 10^{12}$ N\,m$^{-2}$ for (22) Kalliope indicating a range of internal structures from heavily 
fractured or shattered rock to solid rock if all large main-belt binary systems are billions of years old.  If 
a binary system is actually younger than 4.5 Gyr, $\mu Q$ will scale downward by the same factor.

Given the low $\mu Q$ values of (702) Alauda, (45) Eugenia, and (130) Elektra and the implied porosities of the 
primary components based on their densities in Table~\ref{tab:mba}, some of the large main-belt primaries may be 
shattered and porous gravitational aggregates.  In the gravitational-aggregate model of~\citet{gold09}, void 
space (porosity) between the constituent particles, each of which has the rigidity of solid rock of order 
10$^{10}$ N\,m$^{-2}$, and the increased stress at contact points lower the effective rigidity of the aggregate 
as a whole compared to monolithic rock\footnote{\citet{gold09} find that the minimum rigidity of a gravitational
aggregate compares to monolithic rock as 
$\mu \sim \left(8\pi G\rho^{2}R^{2} \mu_{\rm monolith} / 57\epsilon_{\rm Y}\right)^{1/2} \sim 10^{7}\,\left(R/{\rm km}\right) {\rm N\,m^{-2}}$, 
where $\epsilon_{\rm Y} \sim 10^{-2}$ is the yield strain.  A gravitational aggregate 100 km in radius would 
have a minimum rigidity of order 10$^{9}$ N\,m$^{-2}$.} and can account for values of $\mu Q$ of order 10$^{10}$ 
or 10$^{11}$ N\,m$^{-2}$ for gravitational aggregates 100 km in radius and assuming $Q \sim 10$--$100$.  Since 
$\mu Q$ values calculated for (702) Alauda, (45) Eugenia, and (130) Elektra fall below 10$^{10}$ N\,m$^{-2}$, this 
could indicate that another mechanism is assisting tides in the semimajor-axis expansion of these systems, our 
ignorance of which results in a lower $\mu Q$ to account for the work done by the non-tidal process.  If, instead, 
these systems formed near their current configurations, $\mu Q$ would correspondingly increase such that the system 
evolves in semimajor axis more slowly over the same time period.  However, the dependence on initial separation 
is so weak (see Section~\ref{sec:other}) that the systems would have to have formed beyond 10$R_{\rm p}$ to 
raise $\mu Q$ above 10$^{10}$ N\,m$^{-2}$, which is difficult by sub-catastrophic collision.  \citet{durd04} 
find that the material excavated during a sub-catastrophic collision that initially orbits close to the primary 
typically has a semimajor axis of 4--7$R_{\rm p}$, and further collisions among the orbiting material will 
eventually coalesce to form a single secondary orbiting within ``several'' primary radii rather than beyond 
10$R_{\rm p}$. 

The despinning timescales for the binary components of the $R_{\rm s}/R_{\rm p}<0.1$ $(q<0.001)$ systems, 
assuming $\mu Q$ of $1 \times 10^{11}$ N\,m$^{-2}$, are of order 1 Myr for the secondaries (if they began 
their tidal evolution at an initial separation within a handful of primary radii) and much longer than the age 
of the Solar System for the primaries.  Given the relatively short timescales for the secondaries, one would 
expect the secondaries in collisional binaries to have synchronized their rotation rate to their mean motion 
in the mutual orbit even if the systems are much younger than 4.5 Gyr old.  Because the despinning timescales of 
the primaries far exceed the age of the Solar System, however, the fact the systems are still tidally evolving 
does not place a limit on the age of the binaries.  The rapid synchronization of the secondary and the negligible 
despinning of the primary make our ignorance of the initial spin states of the components unimportant for tidal 
evolution of binaries with small secondaries.  In addition, the small size of the secondary with respect to the 
primary and its rapid despinning timescale compared to the assumed age of the binary also make the correction in 
Eq.~\ref{eq:muq} due to accounting for the secondary negligible 
[$\left(R_{\rm s}/R_{\rm p}\right)\left(\tau_{\rm s}/\Delta t\right)\sim10^{-4}$ compared a maximum possible 
contribution of unity].  We also note that the smallest secondaries in this subset are several kilometers in 
diameter, which gives them collisional lifetimes of more than 1 Gyr~\citep{bott05} such that the existence of a 
secondary does not limit the age of a large main-belt binary either.

Taking the median $\mu Q$ of $1 \times 10^{11}$ N\,m$^{-2}$ found for the $q<0.1$ large main-belt binaries, 
equal-mass binary (90) Antiope needs less than 10,000 years to evolve from near-contact to its present 
configuration.  If the components of Antiope are heavily fractured gravitational aggregates, the time needed 
to tidally evolve becomes even shorter by an order of magnitude for every order of magnitude less than $10^{11}$ 
N\,m$^{-2}$ the material strength of Antiope is.  Even for monolithic rock with $\mu Q$ of order $10^{13}$ 
N\,m$^{-2}$, Antiope can fully evolve by tides within 1 Myr.  Though Antiope is a perfect example of the result of 
tidal evolution, as it resides in a stable fully synchronous tidal end state, the system can evolve to its 
observed state so rapidly for any reasonable $\mu Q$ value that one cannot constrain the material properties of 
Antiope well through tidal evolution.  Timescales for (617) Patroclus are longer due to the wide separation of 
its components with tidal despinning acting on million-year timescales assuming the median $\mu Q$ value.  Based 
on its total angular momentum, the Patroclus system must have formed with the components separated by at least 
7$R_{\rm p}$ even if the components once rotated at the cohesionless breakup rate.  However, this wider initial 
separation only affects the calculation of $\mu Q$ by 1\% compared to our standard assumption of 2$R_{\rm p}$.

\subsection{Near-Earth binaries}
\label{sec:nea}

We also consider binary systems in the near-Earth region well-characterized by radar and lightcurve observations, 
all of which have primaries with diameters on the 1-km-scale or smaller.  These systems are most likely the result 
of a spin-up mechanism~\citep{marg02s,rich06,desc08} as evidenced by the $J/J^{\prime}\sim0.4$ values for the 
near-Earth binaries in Table~\ref{tab:nea}.  Since the observational confirmation of the asteroidal YORP 
effect~\citep{tayl07,lowr07,kaas07}, spin-up via anisotropic thermal re-radiation of absorbed 
sunlight~\citep{rubi00,bott06} has become the preferred binary formation mechanism among km-scale parent bodies 
in the near-Earth region and perhaps the main belt~\citep{prav07}.  Further evidence of spin-up comes from the 
rapid rotation of the primaries that, ignoring nearly equal-mass binary (69230) Hermes, spin on average at 
$\sim$90\% of their cohesionless breakup rate with a mean period of 2.8 h and contain over 80\% of the angular 
momentum of the system in their spin.  Though they spin faster than their large main-belt counterparts, the 
primaries of near-Earth binaries contain a smaller fraction of the angular momentum of the system owing to their 
larger secondaries.  The YORP spin-doubling timescale, the time required for an asteroid to change its rotation 
period (or rotation rate) by a factor of two due to radiative torques, is of order 1 Myr for a km-scale asteroid 
in the near-Earth region~\citep{rubi00,vokr02,cape04} allowing the parent body of a near-Earth binary system to
spin-up rapidly and also allowing the post-formation primary component to maintain its rotation near the breakup 
rate. 


\begin{table}[!p]\normalsize
\begin{center}
\begin{tabular}{rlccccccc}
$ $ & Name & $R_{\rm p}$ (km) & $R_{\rm s}/R_{\rm p}$ & $q$ & $\rho$ (g~cm$^{-3}$) & $a/R_{\rm p}$ & $J/J^{\prime}$ & $\mu Q$ (N\,m$^{-2}$)\\
\\
  3671 & Dionysus        & 0.75 & 0.2   & 0.0080 & 2.0 & 5.3 & 0.35 & 7.2\,$\times$\,10$^{7}$\\
  5381 & Sekhmet         & 0.5  & 0.3   & 0.0270 & 2.0 & 3.1 & 0.37 & 3.8\,$\times$\,10$^{9}$\\
 35107 & 1991 VH         & 0.6  & 0.38  & 0.0549 & 2.0 & 6.0 & 0.44 & 1.4\,$\times$\,10$^{8}$\\
 65803 & Didymos         & 0.38 & 0.22  & 0.0106 & 2.0 & 3.0 & 0.41 & 9.8\,$\times$\,10$^{8}$\\
 66063 & 1998 RO$_{1}$   & 0.4  & 0.48  & 0.1106 & 2.0 & 3.6 & 0.47 & 3.8\,$\times$\,10$^{9}$\\
 66391 & 1999 KW$_{4}$   & 0.64 & 0.33  & 0.0359 & 2.0 & 4.0 & 0.37 & 1.6\,$\times$\,10$^{9}$\\
 69230 & Hermes$^{a}$    & 0.3  & 0.9   & 0.7290 & 2.0 & 4.0 & 0.49 & 8.4\,$\times$\,10$^{9}$\\
 85938 & 1999 DJ$_{4}$   & 0.18 & 0.5   & 0.1250 & 2.0 & 4.1 & 0.49 & 3.7\,$\times$\,10$^{8}$\\
 88710 & 2001 SL$_{9}$   & 0.4  & 0.28  & 0.0220 & 2.0 & 3.8 & 0.40 & 5.2\,$\times$\,10$^{8}$\\
137170 & 1999 HF$_{1}$   & 1.75 & 0.23  & 0.0122 & 2.0 & 3.4 & 0.40 & 1.1\,$\times$\,10$^{10}$\\
164121 & 2003 YT$_{1}$   & 0.5  & 0.18  & 0.0058 & 2.0 & 6.4 & 0.40 & 6.6\,$\times$\,10$^{6}$\\
175706 & 1996 FG$_{3}$   & 0.75 & 0.31  & 0.0298 & 2.0 & 3.7 & 0.29 & 2.7\,$\times$\,10$^{9}$\\
185851 & 2000 DP$_{107}$ & 0.4  & 0.41  & 0.0689 & 1.7 & 6.6 & 0.48 & 2.9\,$\times$\,10$^{7}$\\
       & 1994 AW$_{1}$   & 0.5  & 0.49  & 0.1176 & 2.0 & 4.8 & 0.50 & 9.7\,$\times$\,10$^{8}$\\
       & 2000 UG$_{11}$  & 0.13 & 0.58  & 0.1951 & 2.0 & 4.3 & 0.43 & 2.2\,$\times$\,10$^{8}$\\
       & 2002 CE$_{26}$  & 1.73 & 0.09  & 0.0007 & 0.9 & 2.7 & 0.43 & 4.4\,$\times$\,10$^{8}$\\
       & 2004 DC         & 0.17 & 0.21  & 0.0093 & 1.7 & 4.4 & 0.41 & 9.6\,$\times$\,10$^{6}$\\
       & 2005 AB         & 0.95 & 0.24  & 0.0138 & 2.0 & 4.0 & 0.29 & 1.3\,$\times$\,10$^{9}$\\
       & 2005 NB$_{7}$   & 0.25 & 0.4   & 0.0640 & 2.0 & 3.7 & 0.34 & 7.5\,$\times$\,10$^{8}$\\
       & 2006 GY$_{2}$   & 0.2  & 0.2   & 0.0080 & 2.0 & 3.0 & 0.37 & 2.2\,$\times$\,10$^{8}$\\
       & 2007 DT$_{103}$ & 0.15 & 0.33  & 0.0359 & 2.0 & 3.6 & 0.39 & 1.6\,$\times$\,10$^{8}$\\
\end{tabular}
\end{center}
\caption[]{Physical properties of near-Earth binary asteroids with primary radii of order 1 km or smaller.  
Densities are estimated as 2~g\,cm$^{-3}$~\citep[see][for details]{prav07} unless the separation is known 
from radar observations, which allows the density to be calculated via Kepler's Third Law.  $J/J'$ in each 
case roughly satisfies the condition for binary formation via spin-up of a single parent body.  Values of 
$\mu Q$ correspond to tides raised on the primary only acting over the 10 Myr dynamical lifetime of near-Earth 
asteroids.  Accounting for tides raised on the secondary would increase $\mu Q$ by no more than a factor of 
two.  Data are taken from the Ondrejov Asteroid Photometry Project binary asteroid parameters table 
(http://www.asu.cas.cz/$\sim$asteroid/binastdata.htm, 2010 April 8 release) with the exceptions of 2000 
DP$_{107}$ from~\citet{marg02s} and 2004 DC whose parameters are provided by the authors. \\$^{a}$ (69230) 
Hermes is known to be in a fully synchronous state where tidal evolution has ceased.}
\label{tab:nea}
\end{table}

\begin{figure}[!p]
\begin{center}
\includegraphics[angle=90., scale=0.60]{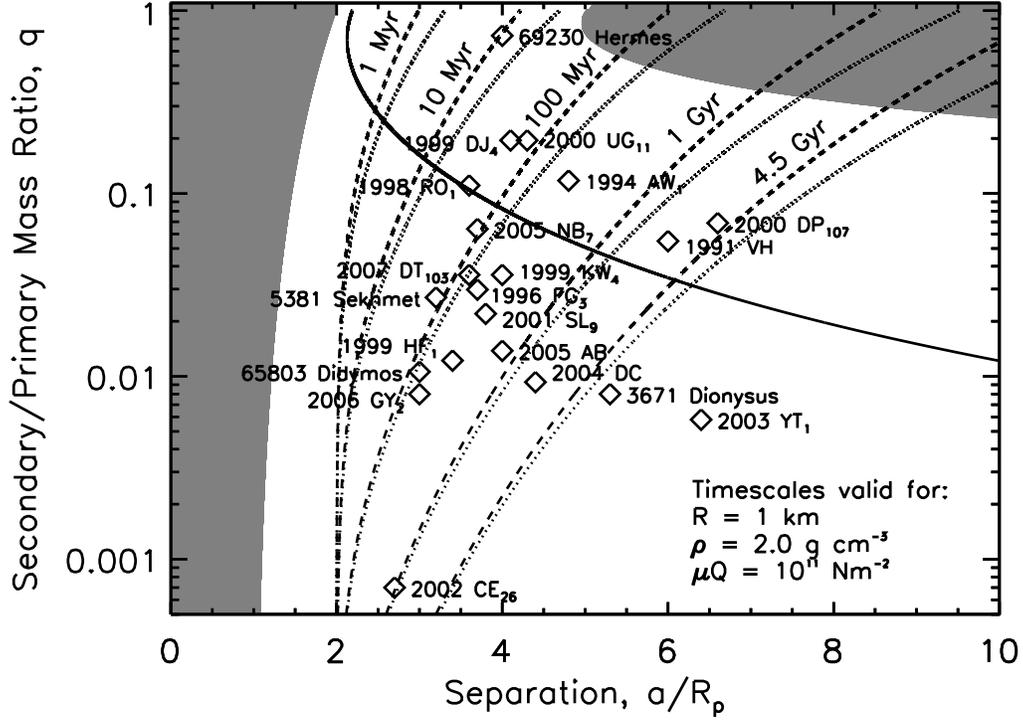}
\caption[]{Mass ratio $q$ and primary-secondary separation $a/R_{\rm p}$ for near-Earth binaries 
along with the angular momentum limit for $J/J^{\prime}=0.5$.  The solid curve is the stability limit. Tidal 
evolution timescales are plotted assuming $\rho_{\rm p, s}=2$ g~cm$^{-3}$, $R_{\rm p}=1$ km, and 
$\mu_{\rm p}Q_{\rm p}=\mu_{\rm s}Q_{\rm s}=10^{11}$ N\,m$^{-2}$ and allowing tides raised on the primary only 
(dashed) or both components (dotted) to contribute over the entire evolution from an initial separation 
of $2R_{\rm p}$.  The spread of the binaries across a range of timescales indicates either a range of ages 
for the binaries, a range of material strengths in terms of $\mu Q$, or a combination of both.}
\label{fig:nea}
\end{center}
\end{figure}

Aside from nearly equal-mass binary (69230) Hermes that has reached its fully synchronous end state~\citep{marg06iau}, 
the binaries in the near-Earth region are also in the midst of a lengthy tidal evolution.  As shown by 
Fig.~\ref{fig:nea}, a handful of systems have evolved past the stability limit, but all remain far from reaching 
the outer synchronous orbit for $J/J^{\prime}=0.4$ that lies at 8$R_{\rm p}$ for $q=0.2$ and rapidly recedes 
further out as $q$ decreases.  Among those beyond the stability limit, (35107) 1991 VH and (185851) 2000 DP$_{107}$ 
have been well studied.  (35107) 1991 VH is believed to be an asynchronous binary as three periods are sometimes 
detected in lightcurves~\citep{prav06}, presumably the mutual orbit period, the rotation of the primary, and the 
nonsynchronous rotation of the secondary.  Tumbling of the primary~\citep{prav06} or the secondary~\citep{marg08dps}
could also explain the third periodicity.  The orbit and rotation periods of the secondary of (185851) 2000 DP$_{107}$ 
suggest synchronization, but the primary has not been despun~\citep{marg02s,prav06}.  These systems illustrate the 
necessary, but not sufficient nature of crossing the stability limit for completing tidal evolution.  Because the 
inner synchronous orbit is buried within the contact limit for the $q\sim0.1$ binaries with $J/J^{\prime}=0.4$ (see 
Fig.~\ref{fig:synch}), the secondaries have tidally evolved outward since the binaries were formed.


The dynamical lifetime of an asteroid in the near-Earth region is of order 10 Myr~\citep{glad97,glad00}, so the 
fact that all near-Earth binary systems lie beyond the generic 10 Myr evolution curve in Fig.~\ref{fig:nea} implies 
either the actual $\mu Q$ values are much weaker than the assumed 10$^{11}$ N\,m$^{-2}$ or that the systems must be 
older than 10 Myr having formed in the main belt prior to injection into the near-Earth region.  Assuming all the 
binaries have tidally evolved for 10 Myr, the median $\mu Q$ in Table~\ref{tab:nea} is 5\,$\times$\,10$^{8}$ 
N\,m$^{-2}$, more than two orders of magnitude smaller than the median $\mu Q$ for the large main-belt binaries.  A 
binary formed in the near-Earth region more recently than 10 Myr ago requires a smaller $\mu Q$ than presented to 
complete the same tidal evolution.  If formed in the main belt, the age of the binaries could extend to, at most, 
the collisional lifetime of a body 1 km in diameter, which is a few hundred million years~\citep{bott05}.  Thus, the 
$\mu Q$ values in Table~\ref{tab:nea} could increase by roughly an order of magnitude if near-Earth binaries formed 
in the main belt.  

For the median $\mu Q$ value of 5\,$\times$\,10$^{8}$ N\,m$^{-2}$, the secondaries in near-Earth binary systems are 
despun on 0.1--1 Myr timescales and thus are expected to be synchronously rotating for systems that are 10 Myr old, 
while the primaries are despun on timescales closer to the dynamical or collisional lifetimes of the bodies.  Nearly 
equal-mass binary (69230) Hermes needs less than 1 Myr to tidally evolve from near-contact to its fully synchronous 
end state assuming the median $\mu Q$ value.  Because the YORP timescale for km-scale asteroids in the near-Earth 
region is similar in magnitude to the tidal despinning timescale for the secondaries, YORP spin-up could hinder, 
negate, or even overcome tidal spin-down, delaying or preventing synchronous lock of the secondaries~\citep{gold09}.  
The YORP timescale is much faster than the tidal despinning timescale for the primaries, which can help the primaries 
retain a rapid rotation (and possibly trigger a mass-lofting scenario) while tides evolve the separation of the 
components.

Though far less rigid than solid rock, $\mu$ as low as 10$^{7}$ N\,m$^{-2}$ for a body 1 km in radius (10$^{6}$ N\,m$^{-2}$
for a body 100 m in radius) remain reasonable within the theory of~\citet{gold09} for gravitational-aggregate structure 
(see Footnote 8).  Assuming $Q \sim 10$--$100$ is applicable, then gravitational aggregates can account for $\mu Q$ 
values of order 10$^{8}$ or 10$^{9}$ N\,m$^{-2}$ for km-scale near-Earth binaries.  Even at such low rigidities, 
$\mu$ still dominates over the stress due to self-gravity $g\rho R$ because of the low surface gravity $g$ and 
small size of the typical near-Earth asteroid.

For binary systems 2004 DC and (164121) 2003 YT$_{1}$ with $\mu Q$ values below 10$^{7}$ N\,m$^{-2}$, the weakness of 
the bodies is difficult to reconcile with Goldreich and Sari's theory of gravitational-aggregate structure.  To 
justify a more rigid structure with $\mu Q$ larger than we calculate in this tidal evolution scenario, one of the 
following must be true:  the system formed recently in nearly its current configuration leaving $\mu Q$ unconstrained, 
the system is older than 10 Myr, or the orbit expansion is aided by another mechanism or mechanisms, \textit{e.g.}, 
BYORP, mass lofting, or close planetary flyby.  Intriguingly, 2004 DC~\citep{tayl08acm} and 2003 YT$_{1}$~\citep{nola04} 
are also both believed to have asynchronous secondaries that are not despun to the mean motion in their mutual orbit 
and have moderate eccentricities greater than 0.1.  These properties place them in the minority among near-Earth binary 
systems that tend to have synchronous secondaries and circularized mutual orbits.  Given that the secondaries are 
rapidly despun by tides compared to orbit expansion timescales in binary systems, finding an asynchronous secondary 
appears to indicate youth, implying that the separation has not expanded much and that these binaries formed in 
essentially their current configurations.  The product $\mu Q$ could be larger by an order of magnitude if the 
binaries are as old as their collisional lifetimes, but then one must explain why the secondaries are not despun 
over the longer timescale.  

The eccentric nature of the mutual orbits of 2004 DC and 2003 YT$_{1}$ could be key to understanding why these systems 
stand out among the rest of the near-Earth binary population, perhaps through entering a tumbling state~\citep{cuk10} 
similar to that of the saturnian satellite Hyperion~\citep{wisd84}.  \citet{cuk10} suggest that after the secondary is 
initially synchronized by tides, outward BYORP evolution can both provide an eccentricity to the mutual orbit and 
trigger episodes of chaotic rotation, breaking the synchronous lock of the secondary.  Without a synchronous secondary, 
BYORP must shut down in these systems, but BYORP may have aided tides in expanding their mutual orbits.  Because of the 
rapid timescales for orbit expansion associated with BYORP, the $\mu Q$ calculated assuming only tidal expansion would 
be artificially lowered due to ignoring the additional BYORP component.  In other words, a smaller $\mu Q$ is needed 
for a rapid orbital expansion, but $\mu Q$ could be larger if tides are not the only orbit expansion mechanism.  
Therefore, these asynchronous, eccentric binaries with very small $\mu Q$ values may be indirect evidence in favor of 
the BYORP scenario. While this scenario would help explain the characteristics of the 2004 DC and 2003 YT$_{1}$ systems, 
one must keep in mind that, contrary to \citet{cuk10}, \citet{mcma10icarus} argue that the BYORP effect should circularize 
the mutual orbit as the separation increases.  Clearly, further research on the dynamical evolution of binary asteroid 
systems due to BYORP is required.  

In addition to BYORP, mass lofting could continuously expand the orbit~\citep{harr09,fahn09} while a close planetary 
encounter could impulsively expand the orbit and provide a moderate eccentricity~\citep{chau95,sche07iau}.  The short
timescale for YORP spin-up could prevent some secondaries from synchronizing and may also be important in keeping the 
primary rotation rapid, hence fueling orbit expansion via mass lofting.  If tidal despinning (and mass lofting) cannot 
regulate the spin of the primary, the primary could conceivably undergo another significant mass-shedding event 
producing a tertiary component leading to triple system, such as (153591) 2001 SN$_{263}$~\citep{nola08} and (136617) 
1994 CC~\citep{broz10dps} in the near-Earth region, or to the ejection of one of the components from the system.  It 
is likely a combination of these effects in the near-Earth region that produce the observed binary configurations as 
well as the seemingly unrealistic values of $\mu Q$ for some systems.

\subsection{Small main-belt binaries}
\label{sec:smba}

Small main-belt binaries, those whose primaries have radii of a few kilometers, represent a population 
that blurs the characteristics of the large main-belt binaries and the near-Earth binaries.  The small 
main-belt binaries share heliocentric orbits in the same region as their much larger cousins, but have 
similar properties to the near-Earth binaries such as angular momentum content.  In Table~\ref{tab:smba}, 
the average $J/J^{\prime}$ is 0.41, similar to the angular momentum required by the spin-fission binary 
formation mechanism, with 73\% of the binaries having $0.3 \le J/J^{\prime} \le 0.5$.  The minority of 
systems have angular momenta consistent both with sub-catastrophic collisions ($J/J^{\prime} < 0.3$) and 
giant impact or capture mechanisms ($J/J^{\prime} > 0.5$).  If formed through spin-up, one would expect 
these systems to have $\mu Q$ values similar to the near-Earth binaries for similar ages.  Because these 
systems are discovered with lightcurve photometry, the known population is limited to systems with size 
ratios of roughly $R_{\rm s}/R_{\rm p} \ge 0.2$, the reliable detection limit for mutual occultation/eclipse 
events~\citep{prav06}.

\thispagestyle{plain}  
\pagestyle{empty}  


\begin{table}[!p]\small
\begin{center}
\begin{tabular}{rlccccccc}
$ $ & Name & $R_{\rm p}$ (km) & $R_{\rm s}/R_{\rm p}$ & $q$ & $\rho$ (g~cm$^{-3}$) & $a/R_{\rm p}$ & $J/J^{\prime}$ & $\mu Q$ (N\,m$^{-2}$)\\
\\
   809 & Lundia          & 3.5 & 0.89 & 0.705 & 2.0 &  4.4 & 0.50 & 6.3\,$\times$\,10$^{13}$\\
   854 & Frostia         & 4.5 & 0.98 & 0.941 & 2.0 &  8.3 & 0.66 & 2.4\,$\times$\,10$^{12}$\\
  1089 & Tama            & 4.7 & 0.90 & 0.729 & 2.0 &  4.6 & 0.51 & 9.0\,$\times$\,10$^{13}$\\
  1139 & Atami           & 2.5 & 0.80 & 0.512 & 2.0 &  6.1 & 0.54 & 2.5\,$\times$\,10$^{12}$\\
  1313 & Berna           & 5.0 & 0.97 & 0.913 & 2.0 &  6.3 & 0.58 & 1.7\,$\times$\,10$^{13}$\\
  1338 & Duponta         & 3.7 & 0.24 & 0.014 & 2.0 &  4.0 & 0.25 & 2.0\,$\times$\,10$^{12}$\\
  1453 & Fennia          & 3.5 & 0.28 & 0.022 & 2.0 &  4.7 & 0.24 & 8.9\,$\times$\,10$^{11}$\\
  1717 & Arlon           & 4.5 & 0.60 & 0.216 & 2.0 & 14.8 & 0.67 & 9.5\,$\times$\,10$^{9}$\\
  1830 & Pogson          & 3.8 & 0.40 & 0.064 & 2.0 &  5.0 & 0.44 & 2.3\,$\times$\,10$^{12}$\\
  2006 & Polonskaya      & 2.8 & 0.23 & 0.012 & 2.0 &  4.2 & 0.31 & 7.3\,$\times$\,10$^{11}$\\
  2044 & Wirt            & 3.5 & 0.25 & 0.016 & 2.0 &  4.2 & 0.27 & 1.4\,$\times$\,10$^{12}$\\
  2131 & Mayall          & 3.7 & 0.30 & 0.027 & 2.0 &  4.8 & 0.39 & 1.1\,$\times$\,10$^{12}$\\
  2478 & Tokai           & 3.5 & 0.86 & 0.636 & 2.0 &  6.1 & 0.56 & 6.6\,$\times$\,10$^{12}$\\
  2577 & Litva           & 2.0 & 0.40 & 0.064 & 2.0 &  6.5 & 0.43 & 1.1\,$\times$\,10$^{11}$\\
  2754 & Efimov          & 3.0 & 0.20 & 0.008 & 2.0 &  3.5 & 0.38 & 1.7\,$\times$\,10$^{12}$\\
  3073 & Kursk           & 2.8 & 0.25 & 0.016 & 2.0 &  7.4 & 0.29 & 2.2\,$\times$\,10$^{10}$\\
  3309 & Brorfelde       & 2.5 & 0.26 & 0.018 & 2.0 &  4.1 & 0.38 & 9.7\,$\times$\,10$^{11}$\\
  3673 & Levy            & 3.2 & 0.27 & 0.020 & 2.0 &  4.6 & 0.36 & 8.7\,$\times$\,10$^{11}$\\
  3703 & Volkonskaya     & 1.4 & 0.40 & 0.064 & 2.0 &  5.0 & 0.37 & 3.1\,$\times$\,10$^{11}$\\
  3749 & Balam           & 2.7 & 0.40 & 0.064 & 2.0 &  6.2 & 0.43 & 2.9\,$\times$\,10$^{11}$\\
  3782 & Celle           & 3.0 & 0.43 & 0.080 & 2.0 &  6.6 & 0.38 & 2.9\,$\times$\,10$^{11}$\\
  3868 & Mendoza         & 4.2 & 0.22 & 0.011 & 2.0 &  4.9 & 0.34 & 5.0\,$\times$\,10$^{11}$\\
  4029 & Bridges         & 4.0 & 0.35 & 0.043 & 2.0 &  3.8 & 0.31 & 9.5\,$\times$\,10$^{12}$\\
  4492 & Debussy         & 5.5 & 0.93 & 0.804 & 2.0 &  6.4 & 0.59 & 1.6\,$\times$\,10$^{13}$\\
  4607 & Seilandfarm     & 4.5 & 0.30 & 0.027 & 2.0 &  5.9 & 0.28 & 4.5\,$\times$\,10$^{11}$\\
  4786 & Titianina       & 3.5 & 0.19 & 0.007 & 2.0 &  4.5 & 0.32 & 3.6\,$\times$\,10$^{11}$\\
  4951 & Iwamoto         & 2.0 & 0.88 & 0.681 & 2.0 & 16.8 & 0.90 & 3.2\,$\times$\,10$^{9}$\\
  5407 & 1992 AX         & 2.0 & 0.20 & 0.008 & 2.0 &  3.3 & 0.36 & 1.1\,$\times$\,10$^{12}$\\
  5477 & 1989 UH$_{2}$   & 1.5 & 0.40 & 0.064 & 2.0 &  5.0 & 0.40 & 3.4\,$\times$\,10$^{11}$\\
  5481 & Kiuchi          & 2.6 & 0.33 & 0.036 & 2.0 &  4.5 & 0.31 & 1.2\,$\times$\,10$^{12}$\\
  5905 & Johnson         & 1.8 & 0.38 & 0.055 & 2.0 &  4.6 & 0.32 & 7.2\,$\times$\,10$^{11}$\\
  6084 & Bascom          & 2.9 & 0.37 & 0.051 & 2.0 &  7.4 & 0.42 & 8.2\,$\times$\,10$^{10}$\\
  6244 & Okamoto         & 2.6 & 0.25 & 0.016 & 2.0 &  4.4 & 0.34 & 6.1\,$\times$\,10$^{11}$\\
  6265 & 1985 TW$_{3}$   & 3.0 & 0.24 & 0.014 & 2.0 &  3.7 & 0.35 & 2.0\,$\times$\,10$^{12}$\\
  6708 & Bobbievaile     & 2.9 & 0.60 & 0.216 & 2.0 &  5.3 & 0.41 & 3.3\,$\times$\,10$^{12}$\\
  7088 & Ishtar          & 0.6 & 0.42 & 0.074 & 2.0 &  4.5 & 0.43 & 1.3\,$\times$\,10$^{11}$\\
  7255 & Huntress        & 2.7 & 0.21 & 0.009 & 2.0 &  3.5 & 0.38 & 1.7\,$\times$\,10$^{12}$\\
  7369 & Gavrilin        & 2.4 & 0.70 & 0.343 & 2.0 &  8.7 & 0.55 & 1.5\,$\times$\,10$^{11}$\\
  8116 & Jeanperrin      & 2.3 & 0.40 & 0.064 & 2.0 &  6.5 & 0.37 & 1.5\,$\times$\,10$^{11}$\\
  8373 & Stephengould    & 3.0 & 0.40 & 0.064 & 2.0 &  6.3 & 0.32 & 3.1\,$\times$\,10$^{11}$\\
  9260 & Edwardolson     & 1.9 & 0.27 & 0.020 & 2.0 &  4.0 & 0.32 & 7.3\,$\times$\,10$^{11}$\\
  9617 & Grahamchapman   & 2.5 & 0.27 & 0.020 & 2.0 &  4.2 & 0.42 & 8.7\,$\times$\,10$^{11}$\\
 10208 & 1997 QN$_{1}$   & 1.7 & 0.46 & 0.097 & 2.0 &  9.1 & 0.47 & 1.4\,$\times$\,10$^{10}$\\
 11264 & Claudiomaccone  & 2.1 & 0.40 & 0.064 & 2.0 &  3.6 & 0.36 & 5.5\,$\times$\,10$^{12}$\\
 15268 & Wendelinefroger & 2.1 & 0.30 & 0.027 & 2.0 &  5.0 & 0.41 & 2.8\,$\times$\,10$^{11}$\\
 16635 & 1993 QO         & 1.9 & 0.35 & 0.043 & 2.0 &  6.0 & 0.48 & 1.1\,$\times$\,10$^{11}$\\
 17260 & 2000 JQ$_{58}$  & 1.6 & 0.26 & 0.018 & 2.0 &  3.5 & 0.31 & 1.0\,$\times$\,10$^{12}$\\
 26471 & 2000 AS$_{152}$ & 3.2 & 0.36 & 0.047 & 2.0 &  6.8 & 0.42 & 1.5\,$\times$\,10$^{11}$\\
 32008 & 2000 HM$_{53}$  & 1.7 & 0.50 & 0.125 & 2.0 &  7.1 & 0.50 & 8.8\,$\times$\,10$^{10}$\\
 34706 & 2001 OP$_{83}$  & 1.6 & 0.28 & 0.022 & 2.0 &  4.4 & 0.38 & 2.9\,$\times$\,10$^{11}$\\
 76818 & 2000 RG$_{79}$  & 1.4 & 0.35 & 0.043 & 2.0 &  3.5 & 0.34 & 2.3\,$\times$\,10$^{12}$\\
\end{tabular}
\end{center}
\caption[]{Physical properties of small main-belt binary asteroids with an average primary radius of 3 km.  
Densities are estimated as 2~g\,cm$^{-3}$~\citep[see][for details]{prav07} in all cases.  The typical $J/J'$ 
roughly satisfies the condition for binary formation via spin-up of a single parent body as with the 
near-Earth binaries.  Values of $\mu Q$ correspond to tides raised on the primary acting over the 1 Gyr 
collisional lifetime of the secondaries.  Accounting for tides raised on the secondary would increase 
$\mu Q$ by no more than a factor of two.  Data are taken from the Ondrejov Asteroid Photometry Project 
binary asteroid parameters table (http://www.asu.cas.cz/$\sim$asteroid/binastdata.htm, 2010 April 
8 release).}
\label{tab:smba}
\end{table}

As seen with binaries like (90) Antiope, systems with nearly equal-mass components can tidally evolve to a 
fully synchronous end state very rapidly.  If we ignore this type of system among the small main-belt 
binaries, where the spin periods have likely lengthened to equal the mutual orbit period, the average spin 
period of the primary components is 3.2 h and makes up $\sim$75\% of the angular momentum of the system.  
This is similar to the 2.8-h average period for primaries in near-Earth binary systems and another indication 
that a similar binary formation process is at work in the near-Earth region and among km-scale main-belt 
asteroids.  The YORP spin-doubling timescale for a parent body a few km in radius orbiting in the main belt is 
of order 100 Myr~\citep{rubi00,vokr02,cape04}.  Thus, in a few hundred million years, a small main-belt asteroid 
could spin up from a mundane period of 6 or 12 h to a period of 3 h or less producing a binary system 
with a rapidly spinning primary, where accounting for the time required to spin up the parent body does not 
drastically reduce the possible age of the resulting binary.  The collisional lifetimes of the km-scale 
secondaries, however, limit the ages of the small main-belt binaries to between roughly 100 Myr and 
1 Gyr~\citep{bott05}.

\begin{figure}[!p]
\begin{center}
\includegraphics[angle=90., scale=0.60]{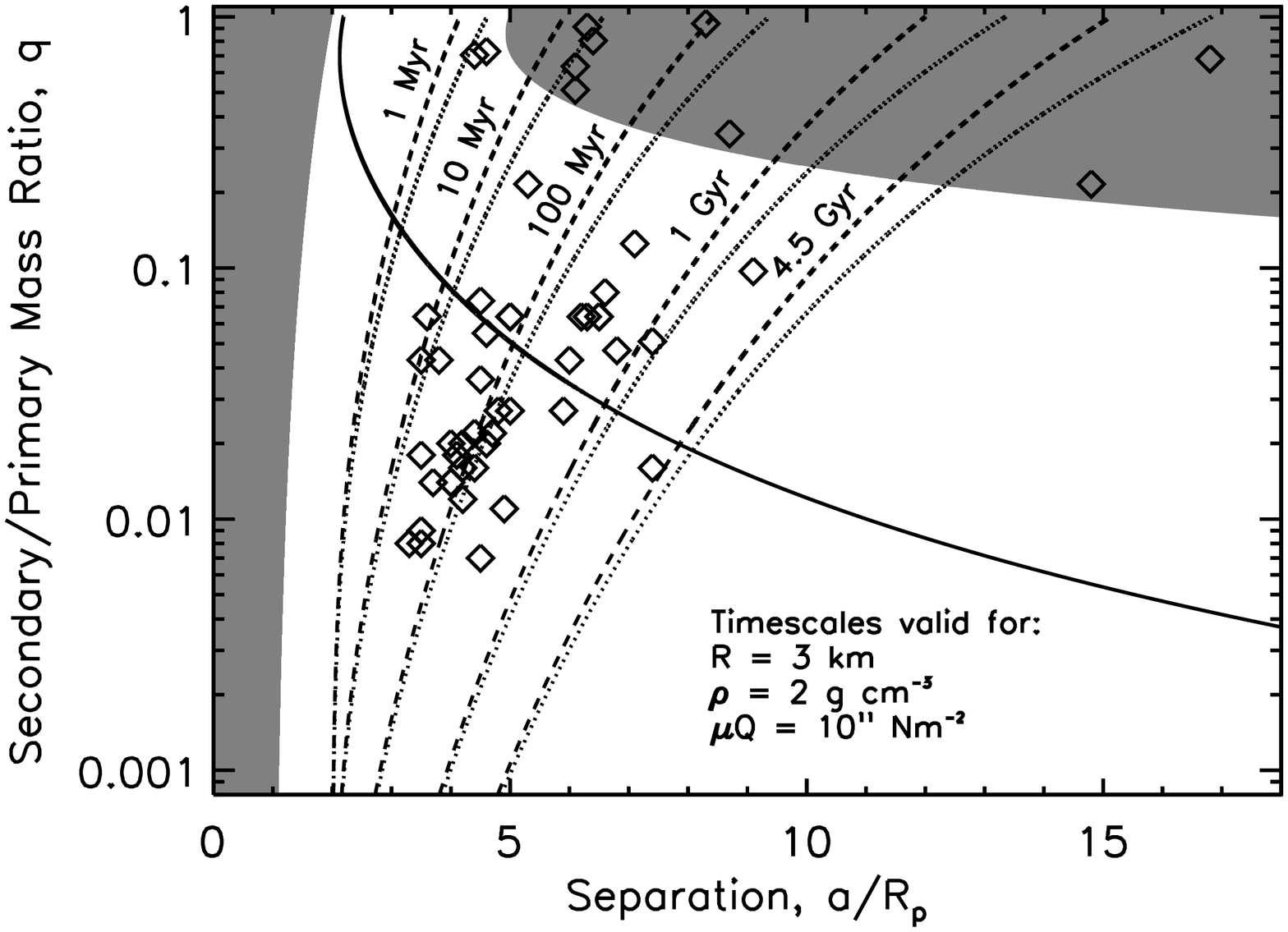}
\caption[]{Mass ratio $q$ and primary-secondary separation $a/R_{\rm p}$ for small main-belt binaries 
along with the angular momentum limit for $J/J^{\prime}=0.5$.  The solid curve is the stability limit. Tidal 
evolution timescales are plotted assuming $\rho_{\rm p, s}=2$ g~cm$^{-3}$, $R_{\rm p}=3$ km, and 
$\mu_{\rm p}Q_{\rm p}=\mu_{\rm s}Q_{\rm s}=10^{11}$ N\,m$^{-2}$ and allowing tides raised on the primary only 
(dashed) or both components (dotted) to contribute over the entire evolution from an initial separation 
of $2R_{\rm p}$.  The binaries tend to cluster near size ratios of 0.2 ($q\sim0.01$), the lower limit for
reliable mutual event detection by photometry, and an age of 100 Myr for the assumed $\mu Q$ value.}
\label{fig:smba}
\end{center}
\end{figure}

Assuming a Phobos-like $\mu Q$ of $10^{11}$ N\,m$^{-2}$, the small main-belt binary systems in Fig.~\ref{fig:smba} 
cluster around an age of 100 Myr with an order-of-magnitude scatter to both younger and older ages.  If small 
main-belt binaries have existed instead for the maximum collisional lifetime of the secondaries of 1 Gyr, the
median $\mu Q$ in Table~\ref{tab:smba} is $7\times10^{11}$ N\,m$^{-2}$.  For this value of the material strength, 
the secondaries in small main-belt binary systems are despun on 1--10 Myr timescales, quicker than the YORP 
timescale, and thus are expected to be synchronously rotating for systems that are 1 Gyr old, while the primaries 
are despun on timescales similar to the collisional lifetimes of the bodies or longer.  Nearly equal-mass 
binaries can despin on 1 Myr timescales assuming the median $\mu Q$ value.

The median $\mu Q$ we calculate is not only larger than that found for the large main-belt binaries, but it is 
reasonable for solid or somewhat fractured rock.  Considering that the majority of small main-belt binaries 
appear to have formed via spin-up, unless the primaries are essentially solid cores left over from a parent 
body that shed its loose material to form a substantial secondary ($R_{\rm s}/R_{\rm p} \ge 0.2$), the median 
$\mu Q$ for evolution over 1 Gyr seems like an overestimate.  If the systems are much younger than their 
collisional lifetimes, for every order of magnitude younger the binary systems are, $\mu Q$ will also reduce 
by an order of magnitude.  If small main-belt binaries are 100 Myr old, the median $\mu Q$ value would be within 
a factor of two of Phobos or a typical large main-belt binary; if small main-belt binaries are 10 Myr old, they 
would have a median $\mu Q$ an order of magnitude stronger than near-Earth binaries of the same age.  An 
order-of-magnitude difference in $\mu Q$ between small main-belt binaries and near-Earth binaries could be a 
size effect since, in Goldreich and Sari's gravitational-aggregate model, the rigidity scales as the radius of 
the body (see Footnote 8).  The average small main-belt primary in Table~\ref{tab:smba} is a factor of five 
larger than the average near-Earth primary in Table~\ref{tab:nea}, and thus could account for the same factor 
in the rigidity of the bodies.  If a size effect cannot account for the difference between the $\mu Q$ values 
found for small main-belt and near-Earth binaries of similar ages, then this discrepancy could indicate that 
the $\mu Q$ values found for near-Earth binaries are artificially lowered due to the presence (and our 
ignorance) of other orbit expansion mechanisms.  

\pagestyle{plain} 


Small main-belt binaries, if formed via spin-up, should form on a rather consistent basis over the last billion 
years, yet we find them to clump together at a similar age in Figure~\ref{fig:smba}.  For the median value of 
$\mu Q$, this would be 1 Gyr, but the median $\mu Q$ value seems at odds with the weaker structure one expects 
for a binary formed via spin-up.  Lowering the age of the population to reduce the median $\mu Q$ of the small 
main-belt binaries would make it seem that the small main-belt binary population is missing an ``older'' component.  
For a Phobos-like $\mu Q$ of $10^{11}$ N\,m$^{-2}$, binaries with an age of about 1 Gyr are under-represented if 
binaries have formed on a consistent basis, and the smaller the median $\mu Q$ of the binaries, the younger the 
observed population and also the larger the proportion of ``missing'' binaries.  If detection bias against longer 
orbit periods in photometric surveys~\citep{prav06} is not the sole cause of the relative dearth of wider binaries 
that would represent an older population for the same $\mu Q$ value, perhaps this is evidence that older (100 Myr 
to 1 Gyr old) binaries have been destroyed via collisions or that another mechanism exists in the main belt that is 
capable of destroying binaries.  BYORP is capable of contracting the mutual orbits of binary systems, leading 
either to the coalescence of the components or a masking of the true age of the system by hindering tidal expansion.  
Unchecked orbit expansion by BYORP could conceivably lead to the loss of the secondary, though complete dissociation 
of the components via BYORP is argued against by~\citet{cuk10}.  The YORP effect is also capable of collapsing 
binaries by removing angular momentum from the system until a fully synchronous orbit no longer 
exists~\citep{tayl09dda}.  Continued YORP spin-up and repeated mass-shedding events could conceivably produce a 
tertiary component in less than the YORP spin-doubling timescale of 100 Myr.  If the triple system is unstable, one 
body could be ejected such that the binaries we observe are simply the most recent incarnation of the system.  Note, 
however, that while the YORP timescale is shorter than the tidal despinning timescale for the primary assuming the 
median $\mu Q$ value, if small main-belt binaries are 10 Myr old instead of 1 Gyr old, $\mu Q$ is reduced, and tidal 
despinning should then overcome YORP spin-up, preventing further mass-shedding events and slowing down the primary's 
rotation from the breakup rate.

\subsection{Other considerations}
\label{sec:other}

Several factors can affect the calculation of $\mu Q$ in addition to the major source of error, the assumed 
age of the binary $\Delta t$.  These include measurement errors in the (1) sizes and densities of the 
components or (2) the current separation $a/R_{\rm p}$, (3) the assumption of an initial separation, and (4) 
neglect of higher-order terms in the tidal potential at small separations.  These considerations are 
discussed in more detail in~\citet{tayl10cmda} and are summarized here:

\begin{enumerate}
\item The calculation of $\mu Q$ scales as 
$\displaystyle{\rho_{\rm p}^{5/2}R_{\rm p}^{2}\,q\left(1+q\right)^{1/2}\propto\rho_{\rm p}^{11/6}\,q\left(1+q\right)^{1/2}}$
for a binary with a known system mass such that underestimates (overestimates) in any of these parameters by 
some percentage lead to underestimates (overestimates) of $\mu Q$ by a larger percentage.  The standard 
assumption of equal component densities could be troublesome, as for (66391) 1999 KW$_{4}$, where the difference 
in component densities~\citep{ostr06} results in a 43\% error in $q$ that can skew $\mu Q$ by a similar amount.
\item An error in the current separation $a/R_{\rm p}$ of 10\% can cause a factor of two error in the 
calculated value of $\mu Q$.  Typical separation uncertainties are estimated as 10--15\% by the Ondrejov
Asteroid Photometry Project.
\item Assuming an initial separation of 2$R_{\rm p}$ does not affect the calculated $\mu Q$ by more than a 
factor of two unless the true initial separation is within 10\% of the current (final) separation.  We note 
in Table~\ref{tab:mba} that the (617) Patroclus binary system must have formed at a wider separation than
the assumed value of 2$R_{\rm p}$, but only if the true initial separation were greater than 12$R_{\rm p}$ 
would the estimate of $\mu Q$ increase by more than a factor of two.
\item Allowing for higher-order terms in tidal potential due to the proximity of the components can increase
$\mu Q$ by 20--30\% for separations near 2$R_{\rm p}$, but the contribution of extra terms quickly falls off 
to 5\% for tidal evolution to 5$R_{\rm p}$ and to 1\% at 10$R_{\rm p}$.
\end{enumerate}

\noindent
None of the above considerations should affect the calculation of $\mu Q$ by more than a factor of two or so 
on their own; only a severe measurement error or several of these factors working in concert would affect an 
order-of-magnitude calculation of the material properties of a binary asteroid system.

Also of concern are solar tides, whose presence are felt through additional tidal distortions raised on
each component.  The ratio of the tidal amplitudes raised on the primary due to the Sun and the secondary 
scales as $\displaystyle{\left(M_{\odot}/M_{\rm s}\right)\left(a/a_{\odot}\right)^3}$, where $M_{\odot}$ 
is the mass of the Sun and $a_{\odot}$ is the heliocentric semimajor axis of the binary system.  For a 
typical near-Earth binary, $M_{\odot}/M_{\rm s}\sim10^{19}$ and $a/a_{\odot}\sim10^{-8}$ such that the 
ratio of tidal amplitudes is $10^{-5}$ rendering solar tides negligible unless the binary has a shallow
perihelion of a few tenths of an astronomical unit~\citep{sche06s,sche07iau}.  Despite their small masses, 
the secondaries in near-Earth binary systems are so close to their primaries that they easily raise the 
dominant tide.  For a typical large main-belt binary $M_{\odot}/M_{\rm s}\sim10^{15}$ and 
$a/a_{\odot}\sim10^{-6}$ such that the ratio of tidal amplitudes is $10^{-3}$, which is still negligible 
in the scope of this discussion since large main-belt asteroids cannot have close flybys of the Sun.  The 
wider separation of main-belt binary components weakens the tide raised by the secondary, but it still 
dominates over the solar tide.  Typical small main-belt binaries have $M_{\odot}/M_{\rm s}\sim10^{17}$ 
and $a/a_{\odot}\sim10^{-8}$ such that the ratio of the tidal amplitudes is $10^{-7}$ making them the 
least affected by solar tides.  This is due to having larger secondaries than the near-Earth binaries and 
having smaller component separations than the large main-belt binaries.  Since the mass of the secondary 
is typically at least two orders of magnitude less than the primary, solar tides on the secondary are also 
negligible.

\begin{figure}[!p]
\begin{center}
\includegraphics[angle=90., scale=0.60]{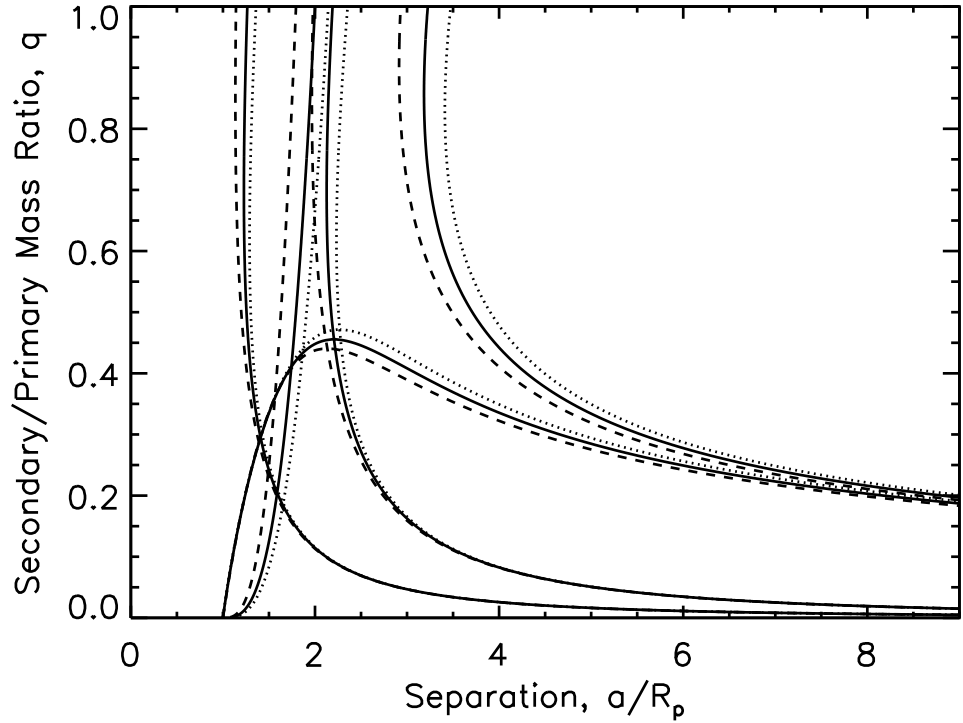}
\caption[]{Fully synchronous orbits (Eq.~\ref{eq:synch}) for binary systems with scaled angular momentum 
$J/J^{\prime}=0.4$ along with the contact limit (Eq.~\ref{eq:contact}), angular momentum limit 
(Eq.~\ref{eq:angmomlimit}), stability limit (Eq.~\ref{eq:stab}), and $E=0$ energy limit (Eq.~\ref{eq:energy}) 
for density ratios of $\rho_{\rm p}/\rho_{\rm s}=0.5$ (dashed), $\rho_{\rm p}/\rho_{\rm s}=1$ (solid), and 
$\rho_{\rm p}/\rho_{\rm s}=1.5$ (dotted).  A less dense primary shifts all curves inward, while a higher-density
primary shifts the curves outward, though the effect is insignificant for mass ratios less than a few tenths.
Spherical shapes with $\alpha_{\rm p, s}=0.4$ are assumed in all cases.}
\label{fig:density}
\end{center}
\end{figure}

Thus far, we have only considered binary components with equal densities.  Applying a density ratio of 
$0.5 \le \rho_{\rm p}/\rho_{\rm s} \le 1.5$ in Eq.~\ref{eq:synch} has little effect on the locations of the 
fully synchronous orbits (see Fig.~\ref{fig:density}) as the density ratio appears only as a pre-factor of the 
mass ratio $q$, which tends to be small.  For typical binaries with $J/J^{\prime}=0.2$ or 0.4 and $q<0.1$, the 
effect of a density disparity on $a_{\rm sync}/R_{\rm p}$ is imperceptible.  For $J/J^{\prime}=0.5$ and $q=1$, 
the effect is almost 10\%, similar to the measurement uncertainty on the separation, though one would expect 
similar-size bodies in a binary system to have similar densities.  Of course, this has assumed that introducing 
a density disparity is the only alteration to Eq.~\ref{eq:synch}, but changing the density ratio will 
necessarily change the mass ratio according to the definition of $q$ (that in turn affects the calculation of 
$\mu Q$ as in point \#1 above) and change the angular momentum $J/J^{\prime}$ as well.  For (66391) 1999 
KW$_{4}$, $\rho_{\rm p}/\rho_{\rm s}=0.7$, which increases the mass ratio $q$ by 43\% and increases $J/J^{\prime}$ 
by 5\% to support the rotation and orbital motion of a more massive secondary.  The stable fully synchronous end 
state shifts from $\sim$120$R_{\rm p}$ to $\sim$70$R_{\rm p}$ due to the combined effects of the larger mass 
ratio (that shifts $a_{\rm sync}/R_{\rm p}$ inward by nearly a factor of two\footnote{In Figs.~\ref{fig:synch} 
and~\ref{fig:density}, for mass ratios of $q<0.5$, the location of the stable fully synchronous orbit rapidly 
recedes such that a small change in mass ratio corresponds to a drastic change in the orbital separation.}) 
and the increased angular momentum of the system (that shifts $a_{\rm sync}/R_{\rm p}$ outward by 10\%).  While 
this change in position is significant, the 1999 KW$_{4}$ system, currently at 4$R_{\rm p}$, will never attain 
the stable fully synchronous end state via tidal evolution due to the insurmountable orbit-expansion timescale 
at wide separations compared to the dynamical lifetime of a near-Earth asteroid and because the stable fully 
synchronous end state lies beyond the Hill sphere of the primary.  It is clearly the effect of the density 
disparity on the mass ratio $q$, capable of changing $q$ by a factor of two for $0.5 \le \rho_{\rm p}/\rho_{\rm s} \le 1.5$, 
that is more important to determining the locations of the stable fully synchronous tidal end state and 
calculating $\mu Q$ than the density ratio as it appears directly in Eq.~\ref{eq:synch}.


We have also only considered spherical bodies with uniform densities thus far.  In Section~\ref{sec:stablimit}, 
we noted how determining the fully synchronous orbits about a nonspherical body requires a more complicated 
treatment of the mutual gravitational potential and the synchronous rotation rate necessary for relative 
equilibrium than presented here.  Making a body ellipsoidal also increases $\alpha$ as it requires more angular 
momentum to rotate an ellipsoid at a given rate about its axis of maximum moment of inertia than it does a 
sphere; on the other hand, concentrating mass toward the center of the body decreases $\alpha$.  For a typical 
binary system with disparate masses, the spin of the primary dominates the angular momentum of the system such 
that a change in $\alpha_{\rm p}$ by some percentage can change the angular momentum by a similar, but lesser, 
percentage and affect the locations of the fully synchronous orbits in Fig.~\ref{fig:synch} accordingly.  For 
systems with disparate masses, the spin of the secondary contributes to the total angular momentum at the 1\% 
level or less such that the exact value of $\alpha_{\rm s}$ is unimportant.  As with the density ratio, the 
$\alpha_{\rm s}/\alpha_{\rm p}$ ratio only appears as a pre-factor of the mass ratio $q$ in Eq.~\ref{eq:synch}, 
and thus is insignificant unless $q \sim 1$; however, equal-mass binaries are dominated by the orbital rather 
than spin angular momentum when residing in a stable fully synchronous orbit.  Therefore, only the value of 
$\alpha_{\rm p}$ is significant and only then for systems with components of unequal mass.  Take, for example, 
a typical near-Earth binary where 80\% of the angular momentum is in the spin of the primary, then allowing 
$\alpha_{\rm p}$ to vary by 25\% about the nominal value of 0.4 would in turn let the angular momentum of the 
system vary by 20\% or between $J/J^{\prime}=0.32$ and 0.48 assuming the nominal angular momentum of 
$J/J^{\prime}=0.4$ for a binary formed by spin-up.  While such a range in $J/J^{\prime}$ opens a wide swath in 
Fig.~\ref{fig:synch}, the stable fully synchronous orbit will typically be far from the primary, ranging from 
tens of primary radii away for $q=0.1$ to hundreds or thousands of primary radii (or more) as $q$ decreases, 
separations that are comparable to or beyond the Hill sphere of the primary.


\section{Discussion}
\label{sec:disc}

We have derived and plotted the locations of the fully synchronous tidal end states for all mass ratios and
angular momenta and find that only binary asteroid systems with nearly equal-mass components such as (90) 
Antiope and (69230) Hermes reside in these end states with all other systems with smaller mass ratios 
undergoing a lengthy tidal evolution process.  Though (90) Antiope and (69230) Hermes are excellent examples 
of the stable fully synchronous tidal end state, these systems can evolve to their current states so rapidly 
for reasonable values of the product of rigidity $\mu$ and tidal dissipation function $Q$ that their material 
properties are difficult to constrain.  Relying instead upon the binary systems with smaller mass ratios, we 
find among 100-km-scale main-belt binaries values of the material strength $\mu Q$ that, for tidal evolution 
over the age of the Solar System, are consistent with solid or fractured rock as one might expect for binaries 
created via sub-catastrophic collisions.  Binaries formed in the near-Earth region via spin-up that have 
evolved over the typical 10 Myr dynamical lifetime require material strengths orders of magnitude smaller than 
their large main-belt counterparts as one might expect for gravitational aggregates of material compared to 
essentially monolithic bodies.  If near-Earth binaries formed in the main belt prior to injection to the 
near-Earth region, the time available for tidal evolution can lengthen and $\mu Q$ can increase in a similar 
manner.  The discovery of many small main-belt binaries with rapidly spinning primaries and angular momenta 
similar to the near-Earth binaries lends some credence to the idea that near-Earth binaries, or some fraction 
of near-Earth binaries, could form in the main belt prior to delivery to the near-Earth region and have binary 
ages that are limited by collisional lifetimes rather than the dynamical lifetime in the near-Earth region.  
An older age for some near-Earth binaries, \textit{i.e.}, 100 Myr versus 10 Myr, would help reconcile some 
especially small values of $\mu Q$ with the model of gravitational-aggregate structure of~\citet{gold09}.  If 
small main-belt binaries are up to 1 Gyr old, the maximum collisional lifetime of their secondaries, their 
material strengths are more consistent with solid rock than gravitational aggregates.  Younger ages similar to 
the near-Earth binaries would reduce $\mu Q$, but leave an older binary population unaccounted for either 
because of detection bias, collisional destruction, or another destruction mechanism.  Smaller $\mu Q$ values 
for the small main-belt binaries would be more consistent with the gravitational-aggregate structure implied 
by the likely formation mechanism of YORP spin-up and would allow tidal spin-down to overcome continued YORP 
spin-up of the primary, which could account for the slightly slower rotation rates of small main-belt primaries 
compared to near-Earth primaries.    

Beyond the question of age, our ignorance of the inherent $Q$ of gravitational aggregates and whether they can 
be described in such idealized terms are significant caveats in this approach to understanding the material 
properties of asteroids.  Especially in the near-Earth region, the assumption that tides are the dominant method 
of mutual-orbit expansion may also be tenuous.  The binary YORP effect, mass lofting, and close planetary 
encounters could conspire to also expand or contract the mutual orbits, either assisting or hindering the 
expansion of the mutual orbit by tides.  A combination of effects altering the semimajor axis would skew the 
resulting $\mu Q$ when considering tides alone; multiple methods of orbital expansion working in concert would 
give the illusion of a small $\mu Q$ value rather than the true material strength of the body.  It is difficult 
to disentangle different orbit-expansion mechanisms on top of not knowing the precise age of a binary system.  
If we take tides to be the dominant orbit-expansion mechanism, then the assumption of the binary age $\Delta t$ 
is easily the largest source of error in the calculation of $\mu Q$ for a binary system causing an 
order-of-magnitude error in $\mu Q$ for an order-of-magnitude error in $\Delta t$.  Other sources of error, 
those summarized in Section~\ref{sec:other} and also from ignoring tides raised on the secondary, will typically 
cause errors of less than a factor of two.  

In our examination of tidal evolution, a few observations have hinted at the presence of a mechanism (or mechanisms) 
in addition to tides that evolves the semimajor axis of binary asteroid systems.  Following~\citet{gold09}, using 
their model for gravitational-aggregate structure and assuming the actual strength of the BYORP torque on the 
mutual orbit is $10^{-3}$ times its maximum possible value,\footnote{A similar factor for the YORP torque was found 
to be $4\,\times\,10^{-4}$ for asteroid (54509) YORP (formerly 2000 PH$_{5}$) by~\citet{tayl07} and~\citet{lowr07}.} 
it is found that BYORP should dominate the semimajor-axis evolution of all near-Earth binary systems with synchronous 
secondaries.  For small main-belt binaries with unequal-mass components, BYORP and tides could contribute comparably 
to the semimajor-axis evolution, and even for large main-belt binaries, BYORP could be important under certain 
circumstances.  Among the near-Earth binaries, the systems most susceptible to BYORP at some point in their 
evolution are 2004 DC and 2003 YT$_{1}$, precisely the systems whose asynchronous secondaries, eccentric mutual 
orbits, and small and worrisome $\mu Q$ values led us to suspect a BYORP component to their evolution in 
Section~\ref{sec:nea}.  Large main-belt binary systems with mass ratios above $\sim$0.01 are dominated by tides at 
their current separations.  Large main-belt binaries with the smallest size ratios are more susceptible to BYORP, 
and we find that those that plot furthest to the right of the 4.5 Gyr curve in Fig.~\ref{fig:mba} and, hence, have 
the smallest calculated $\mu Q$ values: (45) Eugenia, (130) Elektra, and (702) Alauda, are most likely to be 
dominated by the BYORP effect at this time, which would lead to artificially low $\mu Q$ values compared to 
the rest of the binaries in Table~\ref{tab:mba}.  

In the past, at smaller separations, these low mass-ratio systems among the large main-belt binaries would have 
been tide-dominated, which would allow for synchronization of the secondary with a rough transition to 
BYORP-dominated semimajor-axis evolution occurring around 6$R_{\rm p}$ if the strength of the BYORP effect is 
correctly estimated.  Subsequent expansion by BYORP and tides would then produce the currently observed systems 
where assuming tide-dominated evolution finds a smaller value for $\mu Q$ than is necessarily true.  This fact 
is related to point \#3 in Section~\ref{sec:other}:  the calculation of $\mu Q$ is most sensitive to the evolution 
within 10\% of the current separation such that if BYORP is a significant contributor to the semimajor-axis 
expansion in this range, then $\mu Q$ will be skewed to a lower value than is appropriate for the material.  If 
the strength of the BYORP torque is $10^{-5}$ times its maximum possible value instead, the majority of near-Earth 
binaries and the three large main-belt systems mentioned would remain BYORP-dominated in their current 
configurations.  The small main-belt binaries would be tide-dominated currently making it more difficult for 
destruction by BYORP to account for the apparent lack of older systems.  If the strength of the BYORP torque is 
reduced even further to $10^{-7}$ times its maximum possible value, then only 2004 DC and 2003 YT$_{1}$ (and 
possibly 2000 DP$_{107}$) would remain BYORP-dominated.  Given that BYORP could have a significant impact on many 
of the binary systems we have considered, perhaps the best example for the calculation of $\mu Q$ is (22) Kalliope, 
which has a small enough mass ratio that the system is still tidally evolving, unlike (90) Antiope or (69230) 
Hermes, yet not so small that BYORP could be an issue.  Given the slew of sources of uncertainty in the calculation 
of material properties via tidal evolution, Kalliope's $\mu Q$ should lie within a factor of a few of 
$3.3 \times 10^{12}$ N\,m$^{-2}$ based on a 4.5 Gyr evolution, which implies stronger material properties than 
Phobos, but is reasonable for a monolithic or fractured asteroid.

For a tide-dominated binary system, assuming an age equal to the age of the Solar System provides a confident 
order-of-magnitude upper bound on the value of $\mu Q$.  One way to improve on this upper bound is to limit the 
evolution time by using collisional lifetimes (as discussed for near-Earth and small main-belt binaries) or by 
finding binaries in collisional families.  If we assume the binary was formed as the result of the catastrophic 
disruption of a parent body that produced a family of 
asteroids (\textit{i.e.}, EEBs, escaping ejecta binaries, as described by~\citet{durd04}, then the binary 
evolves only over the age of the family.  Changing the evolution time by some factor will affect the value of 
$\mu Q$ in the same manner.  Of the main-belt binaries listed in Tables~\ref{tab:mba} and~\ref{tab:smba}, about 
a dozen (19\%) are believed to be part of the collisional families determined by~\citet{zapp95}, smaller than the 
value of roughly 30\% found among all asteroids~\citep{marz99,bend02}.  Those families the binaries are associated 
with that have age estimates, $\textit{i.e.}$, Flora~\citep{nesv02flora}, Themis~\citep{marz95,marz99,nesv05}, 
and Eos~\citep{vokr06eos}, are believed to be of order 1 Gyr old so that limiting the ages of the binaries to the 
age of the families rather than the age of the Solar System does not change the upper bound of $\mu Q$ by more 
than a factor of a few.  If one could instead identify binary systems among much younger groups, such as the 
Karin or Veritas clusters, both of which are younger than 10 Myr~\citep{nesv02karin,nesv03}, the ages of the 
binaries, and thus their $\mu Q$ values, would be far better constrained.


\section*{Acknowledgments}
\label{sec:ack}

The authors wish to thank Jean-Marc Petit and an anonymous referee for their detailed reviews that improved the 
clarity and quality of the manuscript.  This work was supported by NASA Planetary Astronomy grants NNG04GN31G 
and NNX07AK68G/NNX09AQ68G to Jean-Luc Margot.  


\bibliographystyle{icarus}
\bibliography{TaylorMargot-TidalEvolution}

\end{document}